\documentclass{aa}

\usepackage{amsmath}
\usepackage{graphicx}
\usepackage{float}
\usepackage{subcaption}
\usepackage{appendix}
\usepackage{ulem}
\usepackage{txfonts}
\usepackage{xcolor}
\usepackage[rightcaption]{sidecap}
\usepackage[pdfpagelabels=false]{hyperref}
\hypersetup{colorlinks=true,linkcolor=blue,citecolor=blue,filecolor=blue}

\begin{document}

\title{RedMaPPer Cluster Properties from Two-Dimensional Lensing Shear Maps in the HSC-SSP Survey}

\author{
Chenxu Cui\inst{1,2}  \and
Xiangkun Liu\inst{1,2}  \and
Huanyuan Shan\inst{3,4} \and
Ziwei Li\inst{1,2} \and
Zuhui Fan\inst{1,2}
}

\institute{
South-Western Institute for Astronomy Research (SWIFAR), Yunnan University, Kunming 650500, People's Republic of China \\
\email{liuxk@ynu.edu.cn}
\and
Yunnan Key Laboratory of Survey Science, Yunnan University, Kunming 650500, People's Republic of China
\and
Shanghai Astronomical Observatory (SHAO), Nandan Road 80, Shanghai 200030, People's Republic of China  \\
\email{hyshan@shao.ac.cn}
\and
University of Chinese Academy of Sciences, Beijing 100049, People's Republic of China
}

\offprints{Xiangkun Liu, Huanyuan Shan}

  \abstract
  {Dark matter halos are fundamental structures in the universe and serve as crucial cosmological probes. Key properties of halos—such as their concentration, ellipticity, and mass centroid—encode valuable information about their formation and evolutionary history. In particular, halo concentration reflects the collapse time and internal structure of halos, while measurements of ellipticity and centroid positions provide insights into the shape and dynamical state of halos. Moreover, accurately characterizing these properties is essential for improving mass estimates and for testing models of dark matter. Gravitational lensing, which directly probes the projected mass distribution without relying on assumptions about the dynamical state, has emerged as a powerful observational tool to constrain these halo properties with high precision.}
  {We aim to derive precise constraints on key structural properties of galaxy clusters—including halo concentration, ellipticity, and the position of mass centroids—by directly fitting observed two-dimensional (2D) weak-lensing shear maps with elliptical Navarro–Frenk–White (NFW) models. These measurements help to reveal the internal structure of massive clusters and to quantify systematic uncertainties in stacked lensing analyses.}
  {We perform a 2D weak-lensing analysis of 299 massive clusters selected from the RedMaPPer catalog, using shear measurements from the first-year data release of the Hyper Suprime-Cam Subaru Strategic Program (HSC-SSP). Elliptical NFW profiles are fit to the shear maps with Gaussian priors on halo mass calibrated from the RedMaPPer cluster richness–mass relation. These priors serve to break the mass–concentration degeneracy in the statistical modelling and, to some extent, tighten the constraints on the other parameters of primary interest.}
  {The derived concentration–mass relation exhibits a slightly steeper slope than traditional weak-lensing power-law or upturn models, and agrees more closely with the results from strong lensing selected halos. More massive and lower-redshift clusters tend to have lower concentrations and appear more spherical. The halo ellipticity distribution is characterized by $e = 1 - b/a = 0.530 \pm 0.168$, with a mean of $\langle e \rangle = 0.505 \pm 0.007$. We also detect a bimodal distribution in the offsets between optical centers and mass centroids: some halos are well-aligned with their brightest cluster galaxy (BCG), while others show significant displacements. These results highlight the power of 2D weak-lensing modeling in probing halo morphology and in providing key inputs for understanding and modeling systematic effects in stacked lensing analyses.}
  {}


  \maketitle

\section{Introduction} 
\label{sec:intro}
The formation of large-scale structures in the Universe, as predicted by N-body simulations based on the $\Lambda$ cold dark matter ($\Lambda$CDM) cosmological model, follows a ``bottom-up" process. In this scenario, smaller structures form first and gradually merge into larger ones under the influence of gravity. These hierarchical structures make up the cosmic web \citep{cosmic_web}, where significant concentrations of matter accumulate at the intersections, leading to the formation of massive dark matter (DM) halos. They host the formation of galaxy clusters, with lower-mass halos mainly merging along filamentary structures \citep{1}. Galaxy clusters, the largest self-gravitationally bound systems observed, play a critical role in shaping the formation and evolution of the Universe. Therefore, studying the properties of these clusters would provide crucial insights into the growth of cosmic structures and the nature of dark matter.

The key properties that characterize galaxy clusters, such as the halo mass function  \citep[e.g.,][]{halo_massfunc_press_Schechter,ST1999,Jenkins2001,Tinker2008,Watson2013} and the density profile \citep[e.g.,][]{NFW,4,cM1,2008Duffy,cM3,cM2,18,cM4}, can be used to test and refine our understanding of structure formation and the evolution of large-scale structures, and are thus essential for constraining cosmological models \citep{eROSITA}. A variety of methods have been developed to measure the properties of halos, each with its strengths and limitations. X-ray observations, for example, allow for the measurement of the total mass of clusters based on the temperature and flux profiles of the hot gas assuming spherical symmetry and hydrostatic equilibrium \citep{10}, which can introduce biases in clusters that deviate from spherical shapes and equilibrium. The Sunyaev-Zel’dovich (SZ) effect links the ionized gas in clusters with Cosmic Microwave Background (CMB) observations, and provides crucial, redshift-independent observables that directly probe the integrated electron pressure of the intracluster medium, enabling us to derive fundamental cluster properties like the mass and thermodynamic state of galaxy clusters across cosmic time \citep{11}. However, point sources and foreground signals could contaminate the SZ effect and complicate the measurements \citep{12,2006SZ}. Other methods, such as dynamics-based estimates, focus on galaxy velocity dispersion and distribution \citep{13,14}, but are also sensitive to the galaxy population within clusters \citep{dynamical}.

Weak gravitational lensing (WL) provides an independent and powerful method for studying galaxy clusters. Unlike other techniques, WL does not rely on assumptions about the hydrodynamic state or symmetry of the clusters \citep[e.g.,][]{shan_liu2018,steepness}. The lensing effect arises when the light from background galaxies is distorted due to the gravitational field of massive foreground structures, including galaxy clusters     \citep{16,Hoekstra2008,FuFan2014,GGLreview,Eucild1,Eucild2}. These distortions thus offer a direct way to probe the distribution of dark matter in the clusters and provide valuable information about their mass, shape, and orientation. Recent WL surveys, including Canada–France Hawaii Telescope Lensing Survey \citep[CFHTLenS\footnote{http://cfhtlens.org/};][]{25}, the Dark Energy Survey \citep[DES\footnote{http://www.darkenergysurvey.org/};][]{26}, the Kilo-Degree survey \citep[KiDS\footnote{http://kids.strw.leidenuniv.nl/};][]{KiDS}, the Hyper Suprime-Cam Subaru Strategic Program survey \citep[hereafter the HSC-SSP\footnote{http://hsc.mtk.nao.ac.jp/ssp/} survey;][]{28}, and the Euclid space telescope of the European Space Agency \citep[ESA\footnote{https://www.esa.int//};][]{Eucild2011}, have significantly improved our ability to measure cosmological parameters and the properties of dark matter halos. These surveys make WL a promising tool for addressing long-standing questions about dark matter and the growth of large-scale structures in the Universe.

One specific technique, galaxy–galaxy lensing (GGL), measures the correlation between the positions of foreground galaxies and the shear of background galaxies \citep{GGL1}. This technique provides valuable clues about the relationship between galaxies and their surrounding dark matter halos \citep{GGL2,GGL4,GGL5,GGL3,GGL6,18,17,19,20,21,22,23}. However, systematic errors such as miscentering and non-sphericity still affect the accuracy of mass estimates in weak-lensing studies. Miscentering occurs when the observed center of the cluster differs from the center of the lensing potential (hereafter the lensing center), leading to biases in the measurement of mass and shape \citep{offset_example}. Non-sphericity, which reflects the fact that halos are often triaxial rather than spherical \citep{2}, can also introduce projection effects, which affect mass estimates \citep{2018orientation,2021shan} and offer important insights into the nature of dark matter particles \citep{SIDM_shape}.

To address these challenges, it is essential to employ 2D WL analyses \citep{2DGGL1,2DGGL2,33}, by considering the full two-dimensional shear map. Incorporating realistic cluster halo features like ellipticity and miscentering, 2D WL can provide more robust estimates of individual halo properties compared to traditional stacking methods. This study presents one of the first large-sample applications of 2D WL analysis to investigate the properties of 299 massive RedMaPPer clusters with the shear data from the first year of the HSC-SSP survey, which significantly enhance statistical capabilities. We individually fit the observed shear map of each cluster to the elliptical Navarro-Frenk-White (NFW) \citep{NFW1996,NFW} models. This facilitates measuring key cluster parameters such as concentration, ellipticity, and mass centroid position, revealing new perspectives on the structure and evolution of massive galaxy clusters.

The structure of this paper is as follows.
We describe the observational data, including the lens and source samples in Section \hyperref[sec:2]{\ref{sec:2}}, and introduce the methodology and fitting procedure in Section \hyperref[sec:3]{\ref{sec:3}}. In Section \hyperref[sec:4]{\ref{sec:4}}, we show our results. Finally, conclusions are presented in Section \hyperref[sec:5]{\ref{sec:5}}. Throughout the paper, we assume a flat universe with the matter density $\Omega_{\mathrm{m}}=1-\Omega_{\mathrm{\Lambda}}=0.315$, the dimensionless Hubble constant $h=0.674$, the baryon density $\Omega_\mathrm{b}h^2=0.0224$, the spectral index $n_\mathrm{s}=0.965$, and the normalization of the matter power spectrum $\sigma_8=0.811$ \citep{Planck2018}. Unless otherwise stated, we quote parameter constraints using the median of the posterior distribution as the best-fit value, with uncertainties corresponding to the 68\% credible interval.

\section{Data} \label{sec:2}

\subsection{Cluster sample} \label{subsec:2.1}

In this paper, we study the properties of clusters from the Red-sequence Matched-filter Probabilistic Percolation (RedMaPPer) cluster finding algorithm \citep[version 6.3,][]{redMaPPer}, found in the Sloan Digital Sky Survey (SDSS) Data Release 8 \citep[DR8,][]{DR8} catalog. This catalog covers a 10,000 $\deg^2$ sky area and contains $26,111$ galaxy clusters over the redshift range z $\in$ [0.08, 0.55], each assigned a richness parameter $\lambda$. The parameter $\lambda$ is the sum of the membership probabilities of galaxies within a richness-scaling radius $R_\lambda=(1.0 h^{-1} \mathrm{Mpc})(\lambda/100.0)^{0.2}$, which is proportional to halo mass. 

We select clusters with spectroscopic redshift measurements to ensure precise determination of their positions along the lines of sight and require that the richness $\lambda$ $>$ 20, yielding a sample of $15,657$ massive clusters. For each RedMaPPer cluster, five galaxy candidates are assigned probabilities $P_{\mathrm{cen}}$, indicating their likelihood of being the optical center galaxy. We then designate the galaxy with the highest probability as the observed center. The displacement between this observed center and the center determined by weak lensing is referred to as the BCG offset. For a full description of the RedMaPPer v6.3 SDSS DR8 catalog, refer to \cite{RedMaPPerCat}.

\subsection{Weak-lensing data} \label{subsec:2.2}
The HSC-SSP is an extensive, wide-field multi-band imaging survey designed to tackle a broad range of scientific inquiries, spanning from cosmology to solar system bodies \citep{28}. A key objective of this survey is the weak lensing study of wide-layer observations, targeting a sky coverage of approximately $1,400\deg^2$. The wide-layer data in the latest third data release has achieved a full depth of $\sim 26$ mag at $5\sigma$ across all five bands (grizy) for an area of about $670\deg^2$ \citep{HSCSSPDR3}. The three-year shear catalog has also been completed \citep{HSCSHEARDR3}, but was only made publicly available very recently.

Therefore, in this paper, we utilize the first-year HSC-SSP shear catalog (S16A) \citep{HSC}. The shapes of galaxies are obtained using the re-Gaussianization point spread function (PSF) correction method on the $i$-band coadded images \citep{HSC_shear}. To ensure the robustness of shear measurements and exclude spurious detections as well as blends that may introduce contamination, we retain only those galaxies that meet the selection criteria specified in Table 4 of \cite{HSC}. The photometric redshift (photo-z) for each galaxy is determined on the basis of HSC five-band photometry, and the specific catalog derived from the Direct Empirical Photometric code \citep[DEMP;][]{DEMP} is used in our study \citep{photozHSCDR1}. It has been demonstrated to exhibit high accuracy, with a scatter of $\sigma[\Delta z_\mathrm{p}/(1 + z_\mathrm{p})]\sim 0.05$ and an outlier rate of $\sim 15$ percent for galaxies in the redshift range of $0.2 \le z_\mathrm{p} \le 1.5$, where $z_\mathrm{p}$ is the optimal photo-z value derived from its probability distribution for a given galaxy. Here, we adopt $z_\mathrm{p}$ as the redshift estimate $z_\mathrm{s}$ for each source galaxy, and then select the lens (cluster)-source pairs that satisfy the criterion $\Delta z=z_\mathrm{s}-z_\mathrm{l}>0.2$, where $z_\mathrm{l}=z_\mathrm{cl}$ denotes the spectroscopic redshift of the cluster studied. This criterion ensures a clear separation between the foreground cluster and background galaxies. Furthermore, to ensure an adequate signal-to-noise ratio, we require the average number density of source galaxies to be $\ge 10/\mathrm{arcmin}^2$, within a circular region centered on the observed center of the foreground cluster and extending to a radius of $0.5\deg$.

\begin{figure}[t]
\centering
\includegraphics[width=\linewidth, angle=0]{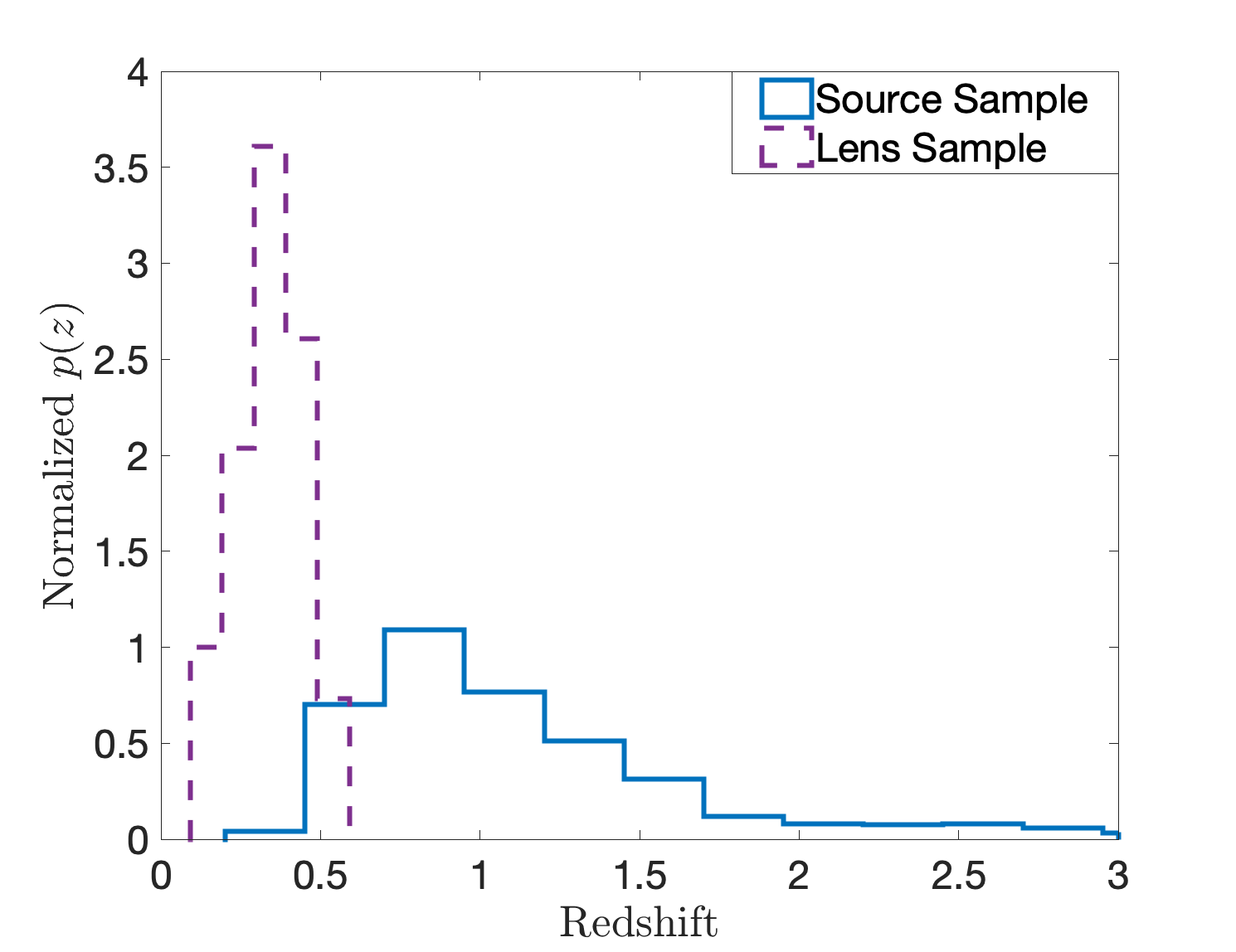}
\caption{Normalized distribution of spectroscopic redshifts for the lens sample (purple dashed) and photometric redshifts for the source sample (blue solid).}
\label{fig hist}
\end{figure}

After cross-matching the RedMaPPer clusters with the sky coverage of the HSC-SSP survey and applying all the aforementioned selection criteria, we ultimately obtain a sample consisting of 299 lens (cluster)-source pairs. The normalized redshift distributions of the entire lens and source samples are presented in Fig.\hyperref[fig hist]{\ref{fig hist}}.

\section{Methods} \label{sec:3}
\subsection{Lensing signal} \label{subsec:3.1}

The weak-lensing effect can be characterized by the second derivatives of the lensing potential $\psi$ via the Jacobian matrix $\mathcal{A}$ \citep{16}:

\begin{equation}
\mathcal{A} = \left(\delta_{ij}-\frac{\partial^2\psi(\boldsymbol{\theta})}{\partial \theta_i\partial\theta_j}\right)=\begin{pmatrix}
1 - \kappa - \gamma_1 & -\gamma_2 \\
-\gamma_2 & 1 - \kappa + \gamma_1
\end{pmatrix},
\end{equation}
where the convergence $\kappa$ and lensing shear $\boldsymbol{\gamma}$ in the complex form of $\gamma_1+i\gamma_2$, respectively, induce an isotropic size change and elliptical shape distortion on the observed image of the background galaxy. Both of them are related to the second derivatives of the lensing potential, and are thus not independent of each other.

The convergence $\kappa$, specifically, is directly related to the projection of line-of-sight density fluctuations weighted by the lensing efficiency factor, which is expressed as the ratio $\Sigma/\Sigma_\mathrm{crit}$, with $\Sigma$ indicating the projected mass density of the lens and $\Sigma_{\mathrm{crit}}$ denoting the critical mass density.

\begin{equation}
\Sigma_{\mathrm{crit}} = \frac{c^{2}}{4{\pi}G} \frac{D_\mathrm{s}}{D_\mathrm{l} D_\mathrm{ls}},
\label{eq:kappa}
\end{equation}
where $D_\mathrm{s}$, $D_\mathrm{l}$, and $D_\mathrm{ls}$ are the angular diameter distances from the observer to the source, from the observer to the lens, and from the lens to the source, respectively. $c$ denotes the speed of light, and $G$ stands for the Newtonian gravitational constant. As an indicator of lensing efficiency, the critical mass density $\Sigma_{\mathrm{crit}}$ for each galaxy cluster is calculated from the redshift distribution of its background sources.
 
On the other hand, the observables for a weak-lensing survey are often the ellipticities of galaxies, which encompass both the intrinsic shape noise of the galaxies and the weak lensing shears $\boldsymbol{\gamma}$, or more precisely, the reduced shears $\boldsymbol{g}=\boldsymbol{\gamma}/(1-\kappa)$ \citep{SS1997}.

The first-year HSC-SSP shear catalog was produced on the coadded $i$-band images using the re-Gaussianization PSF correction method \citep{HSC_shear,HSC}. The core idea of estimating galaxy shape with this method is to apply a Gaussian profile featuring elliptical isophotes to fit the image, and the components of the distortion are defined as follows:

\begin{equation}
\left( e_1, e_2 \right) = \frac{1 - \left( b/a \right)^2}{1 + \left( b/a \right)^2} \left( \cos 2\phi, \sin 2\phi \right),
\label{eq:e12}
\end{equation}
where $b/a$ is the axis ratio and $\phi$ is the position angle of the major axis with respect to the equatorial coordinate system (sky coordinates). The ensemble average distortion is then an estimator for the reduced shear $\boldsymbol{g}$,

\begin{equation}
\left(\hat{g}_{1}, \hat{g}_{2}\right) = \frac{1}{2\mathcal{R}} \langle (e_{1}, e_{2}) \rangle,
\end{equation}
where $\mathcal{R}$ is known as the ``shear responsivity" and represents  the response of the distortion to a small shear \citep{Kaiser1995,BJ2002,shear_R}. $\mathcal{R}\approx 1 - e^{2}_\mathrm{rms}$, where $e_\mathrm{rms}$ is the rms intrinsic distortion per component, typically $e_\mathrm{rms} \simeq 0.4$.

To carry out comprehensive analyses of our cluster sample, we work directly on the 2D distortion maps. For all the clusters, we adopt a pixel size of $1\times 1~\mathrm{arcmin}^2$ centered on the observed center (the position of the BCG), and the lensing distortion is measured in a statistical sense, i.e., by averaging the galaxy shapes over an arcminute scale to mitigate the impact of dominant intrinsic shape noise \citep{33}. Due to their varying masses and redshifts, different clusters occupy different angular sizes on the sky. Therefore, for each cluster, we give a first estimate of its mass $M_\mathrm{est}$ using the richness-mass relation provided in \cite{mass-richness_equation}, calculate the cosmic matter density $\rho_\mathrm{m}(z_\mathrm{cl})$ at the cluster's redshift $z_\mathrm{cl}$, and derive an estimate of halo radius $r_{200\mathrm{m}}=\sqrt[3]{3M_\mathrm{est}/{(4\pi\cdot200\rho_\mathrm{m}(z_\mathrm{cl}))}}$, which is defined as the radius within which the mean halo density is 200 times the cosmic matter density $\rho_\mathrm{m}(z_\mathrm{cl})$. The total number of pixels on each side could be subsequently obtained by rounding up the angular size $\theta_\mathrm{cl}=2\times r_{200\mathrm{m}}/D_A(z_\mathrm{cl})$ of each cluster in units of arcminute, where $D_A(z_\mathrm{cl})$ is the angular diameter distance of the cluster. Pixels without any background galaxies are treated as masked and are excluded from the fitting process; these regions naturally appear blank in the shear map (e.g., Fig.~\ref{fit_results_onecluster})

We follow the approach outlined in \cite{HSC} to compute the weighted shear responsivity factor $\hat{\mathcal{R}}$, utilizing the per-object estimates of rms distortion $e_{\mathrm{rms},i}$ within each $1\times 1~\mathrm{arcmin}^2$ pixel and the inverse variance weights $w_i$ which encode the shape-measurement uncertainty and intrinsic ellipticity dispersion for each galaxy. Specifically, the shear responsivity factor $\hat{\mathcal{R}}(\boldsymbol{\theta}_l)$ in the $l-$th pixel (its angular position $\boldsymbol{\theta}_l$) is given by

\begin{equation}
\hat{\mathcal{R}}(\boldsymbol{\theta}_l) = 1 - \frac{\sum_{i\in\boldsymbol{\theta}_l} w_i e_{\text{rms}, i}^2}{\sum_{i\in\boldsymbol{\theta}_l} w_i},
\end{equation}
where $i\in\boldsymbol{\theta}_l$ indicates that the summation runs over source galaxies contained within the $l-$th pixel.

Similarly, the ensemble estimates for calibration biases in the $l-$th pixel, consisting of both the multiplicative term $\hat{m}(\boldsymbol{\theta}_l)$ and additive term $\hat{c_\alpha}(\boldsymbol{\theta}_l)$, are also derived as a weighted sum of their respective catalog estimates $m_i$ and $c_{\alpha,i}$ within the $l-$th pixel,

\begin{equation}
\hat{m}(\boldsymbol{\theta}_l) = \frac{\sum_{i\in\boldsymbol{\theta}_l} w_i m_i}{\sum_{i\in\boldsymbol{\theta}_l} w_i},
\end{equation}

\begin{equation}
\hat{c_\alpha}(\boldsymbol{\theta}_l) = \frac{\sum_{i\in\boldsymbol{\theta}_l} w_i c_{\alpha,i}}{\sum_{i\in\boldsymbol{\theta}_l} w_i},
\end{equation}
where $\alpha=1,2$, denoting the two components of distortion throughout the paper.

Finally, the reduced shear in the $l-$th pixel is estimated as \citep{33,HSC}

\begin{equation}
g_\alpha(\boldsymbol{\theta}_l)=\frac{\sum_{i\in\boldsymbol{\theta}_l} w_i e_{\alpha,i}}{2\mathcal{R}(\boldsymbol{\theta}_l)\left(1+\hat{m}(\boldsymbol{\theta}_l)\right) \sum_{i\in\boldsymbol{\theta}_l} w_i}-\frac{\hat{c_\alpha}(\boldsymbol{\theta}_l)}{1+\hat{m}(\boldsymbol{\theta}_l)}.
\label{eq:shear}
\end{equation}

For the angular scales of interest, the weak-lensing approximation holds, with $g_\alpha\approx\gamma_\alpha$. In subsequent analyses, this approximation is used only when constructing the large-scale-structure covariance, see Eq.~(\ref{eq:Clss}); both the model and the data vectors are expressed in terms of the reduced shear $g_\alpha$.

To estimate shape noise per pixel, we randomly rotate source galaxies to eliminate correlated lensing signals and apply the same procedures described above to generate 2D noise distortion maps. For each cluster, we perform 1000 random rotations and consequently obtain 1000 noise distortion maps. The shape noise $\sigma_\gamma(\boldsymbol{\theta}_l;\alpha)$, also comprising two components, is then determined by calculating the standard deviation for each pixel across these 1000 noise distortion maps. 

\subsection{Cluster mass model} \label{subsec:3.2}
Clusters in the collisionless cold dark matter universe typically exhibit a highly non-spherical structure that is well fitted by a triaxial density profile \citep[e.g.,][]{2}. We thus construct an elliptical lens model \citep{33} by incorporating an ellipticity into the isodensity contour, under the assumption that the cluster mass distribution can be described by a single halo component whose radial distribution follows NFW density profile. Such an elliptical model provides a better description of halos compared to the spherical one \citep{2,KE2005,Allgood2006}, since the projection of a triaxial halo along any direction results in elliptical isodensity contours on the convergence map \citep{Oguri2003,Oguri2004,2404.00169}. Specifically, the following mass model is adopted in the analyses:

\begin{equation}
\kappa(x,y) = \kappa_\mathrm{sph}(\eta),
\end{equation}

\begin{equation}
\eta^{2} =x'\hspace{0.1em}^2 + \frac{y'\hspace{0.1em}^2}{(1-e)^2},
\end{equation}

\begin{equation}
x' = x \cos\theta_e + y \sin \theta_e,
\end{equation}

\begin{equation}
y' = -x \sin\theta_e + y \cos \theta_e,
\end{equation}
where $\kappa_\mathrm{sph}(\eta)$ denotes the radial convergence for the spherical NFW model.  The halo ellipticity $e$ is defined as $e=1-b/a$, where $a$ and $b$ are the major and minor axis lengths of the isodensity contour. Here we adopt a coordinate system where the coordinate origin set at the observed center of the cluster $(x_c,y_c)$ and position angle $\theta_e$ measured counterclockwise from the positive $x$-axis.

To calculate $\kappa_\mathrm{sph}(\eta)$, we adopt the NFW density profile \citep{NFW1996,NFW}, which is given by

\begin{equation}
\rho(r) = \frac{\rho_s}{\left(r/r_s\right) \left(1 + r/r_s\right)^2},
\end{equation}
where $\rho_s$ and $r_s$ are the characteristic mass density and scale radius, respectively. It is completely characterized by two parameters, the mass and the halo concentration. The corresponding convergence $\kappa_\mathrm{sph}(\eta)=(2\rho_s r_s/\Sigma_\mathrm{crit})f(x=\eta/r_s)$ with \citep{lensing_method}

\begin{equation}
f(x)=\left\{\begin{array}{ll}
\frac{1}{x^{2}-1}\left[1-\frac{2}{\sqrt{1-x^{2}}} \operatorname{arctanh} \sqrt{\frac{1-x}{1+x}}\right] & x<1 \\
\frac{1}{3} & x=1 \\
\frac{1}{x^{2}-1}\left[1-\frac{2}{\sqrt{x^{2}-1}} \operatorname{arctan} \sqrt{\frac{x-1}{1+x}}\right] & x>1
\end{array}\right..
\end{equation}

To compute the lensing efficiency, we use the spectroscopic redshift $z_\mathrm{cl}$ of each cluster and the point-estimate photometric redshifts $z_{\mathrm{p},i}$ of background galaxies to calculate the critical surface mass density $\Sigma_{\mathrm{crit},i}$ for each source galaxy, according to Eq.~(\ref{eq:kappa}). The effective critical surface density is then obtained by averaging over all selected background sources within the cluster region using lensing weights $w_i$:

\begin{equation}
\Sigma_\mathrm{crit,eff} =  \frac{\sum_{i\in \boldsymbol{\theta}_\mathrm{cl}} w_i \Sigma_{\mathrm{crit},i}}{\sum_{i\in \boldsymbol{\theta}_\mathrm{cl}} w_i},
\end{equation}
where the summation is over all background galaxies within a $\theta_{\mathrm{cl}}\times\theta_{\mathrm{cl}}$ region of the cluster center.

The lensing shear for the elliptical mass model is computed from $\kappa(x,y)$ using the relationship between the convergence $\kappa$ and the shear $\boldsymbol{\gamma}$ \citep{KS1,KS2,KS3}. In particular, we use their relation in the Fourier space with

\begin{equation}
\hat{\boldsymbol{\gamma}}(\boldsymbol{k}) = \pi^{-1} \hat{D}(\boldsymbol{k})\hat\kappa(\boldsymbol{k}),
\label{eq:gamma}
\end{equation}
and
\begin{equation}
\hat{D}(\boldsymbol{k})=\pi \frac{k^{2}_{1}-k^{2}_{2}+2ik_{1}k_{2}}{k^{2}_{1} + k^{2}_{2}}.
\end{equation}
The reduced shear is then obtained as $\boldsymbol{g}=\boldsymbol{\gamma}/(1-\kappa)$.

In summary, the elliptical mass model is specified by 6 parameters:

\begin{equation}
\boldsymbol{p} \equiv \{M_{200\mathrm{m}},c_{200\mathrm{m}},e,\Delta x,\Delta y,\theta_e\},
\label{eq:gamma}
\end{equation}
where $M_{200\mathrm{m}}$ is the mass within the halo radius $r_{200\mathrm{m}}$, $c_{200\mathrm{m}}$ is the concentration parameter defined as the ratio $r_{200\mathrm{m}}/r_s$, $e$ is the halo ellipticity, $\Delta x$ and $\Delta y$ represent the offsets along the $x$- and $y$-axes, respectively, from the observed cluster center $(x_c,y_c)$, and $\theta_e$ is the position angle.

\subsection{2D weak-lensing fitting} \label{subsec:3.3}
To constrain cluster properties, we compare the pixelized 2D distortion field $g_\alpha(\boldsymbol{\theta}_l)$, as described in Section \ref{subsec:3.1}, with the reduced shear $g_\alpha^\mathrm{m}(\boldsymbol{\theta}_l;\boldsymbol{p})$ predicted by the elliptical mass model, where $\boldsymbol{p}$ denotes the parameter set defined in Section \ref{subsec:3.2}. Specifically, we calculate the $\chi^2$ as follows:

\begin{equation}
\begin{aligned}
\chi^2=&\sum_{\alpha,\beta=1}^{2}\sum_{l,n=1}^{{N_\mathrm{pix}}}\left[g_\alpha(\boldsymbol{\theta}_l)-g_\alpha^\mathrm{m}(\boldsymbol{\theta}_l;\boldsymbol{p})\right]\left[\boldsymbol{C}^{-1}\right]_{\alpha\beta,ln}\\
&\times \left[g_\beta(\boldsymbol{\theta}_n)g_\beta^\mathrm{m}(\boldsymbol{\theta}_n;\boldsymbol{p})\right],
\end{aligned}
\end{equation}
where the indices $\alpha$ and $\beta$ run over the two components of reduced shear and the indices $l$ and $n$ denote the position of the pixels. $\boldsymbol{C}$ is the error covariance matrix and $\boldsymbol{C}^{-1}$ is its inverse. It is expressed as 

\begin{equation}
\boldsymbol{C}=\boldsymbol{C}^\mathrm{shape} + \boldsymbol{C}^\mathrm{lss}.
\end{equation}

The term $\boldsymbol{C}^\mathrm{shape}$ is dominated by intrinsic shape noise and measurement error. This noise term is expected to be uncorrelated between different pixels; therefore, it is diagonal and is given by

\begin{equation}
\left[\boldsymbol{C}^\mathrm{shape}\right]_{\alpha\beta,ln}=\delta_{\alpha\beta}^K\delta_{ln}^K \sigma^2_\gamma(\boldsymbol{\theta}_l;\alpha),
\end{equation}
where $\delta_{\alpha\beta}^K$ and $\delta_{ln}^K$ represent the Kronecker delta function, and $\sigma_\gamma(\boldsymbol{\theta}_l;\alpha)$ denotes the shape noise estimated from the noise distortion maps, as described in Section \ref{subsec:3.1}.

The term $\boldsymbol{C}^\mathrm{lss}$ presents the covariance due to the large-scale structure, which is given in \citet{Hoekstra2003}.

\begin{equation}
\left[\boldsymbol{C}^\mathrm{lss}\right]_{\alpha\beta,ln} = \xi_{\alpha\beta}(|\theta_l - \theta_n|),
\label{eq:Clss}
\end{equation}
the cosmic shear correlation function, \(\xi_{\alpha\beta}\), is assumed to depend only on the length of the vector connecting the two points \(\theta_l\) and \(\theta_n\), due to the statistical isotropy of the Universe. Specifically, the shear correlation functions $\xi_{\pm}$ are defined through combinations of the two shear components as follows:

\begin{align}
\xi_{11}(r) &= (\cos 2\phi)^2 \, \xi_{tt}(r) + (\sin 2\phi)^2 \, \xi_{\times\times}(r),\\
\xi_{22}(r) &= (\sin 2\phi)^2 \, \xi_{tt}(r) + (\cos 2\phi)^2 \, \xi_{\times\times}(r),\\
\xi_{12}(r) &= \cos 2\phi \sin 2\phi \left[ \xi_{tt}(r) - \xi_{\times\times}(r) \right],
\end{align}
here, $ r = |\theta_l - \theta_n| $, and $\phi$ represents the position angle between the coordinate $x$-axis and the vector $\theta_l - \theta_n $. The functions $\xi_{tt} $ and $\xi_{\times\times}$ correspond to the tangential and cross-component shear correlation functions, respectively \citep{16}. Both of the functions are calculated by the \texttt{pyccl} package\footnote{https://github.com/LSSTDESC/CCLX} \citep{pyccl}.

We utilize the Markov Chain Monte Carlo (MCMC) sampler, \texttt{emcee} package\footnote{https://emcee.readthedocs.io/en/stable} \citep{emcee}, to explore the $\chi^2$ surface, employing a standard Metropolis-Hastings sampling method with a multivariate Gaussian proposal distribution. 

Among the six parameters in our model, we impose a narrow Gaussian prior on the halo mass, with the mean value derived from the mass–richness relation presented in \citet{mass-richness_equation},

\begin{equation}
\ln\left(\frac{M_{200m}}{h^{-1}_{70}{~10^{14}} M_{\odot}}\right) = A + B\times \ln\left(\frac{\lambda}{60}\right),
\end{equation}
where $A=1.72$ and $B=1.08$. The cluster richness $\lambda$ and its measurement uncertainty $\sigma_\lambda$ are taken from the RedMaPPer catalog, and the corresponding mass uncertainty $\sigma_M$ is obtained by propagating errors through the above relation. This allows us to construct a cluster-specific Gaussian prior on halo mass. Our primary interest lies in constraining the structural properties of galaxy clusters, including halo concentration, ellipticity, and the offset between the WL mass centroid and the BCG position. However, uncertainties in halo mass can propagate into these structural parameters and bias their estimation — for example, a massive halo with a downward fluctuation in mass may be incorrectly assigned to a lower mass bin, and vice versa \citep{Duwei2014}. Incorporating a mass prior thus helps reduce such biases and improves the robustness of our inference on the connection between halo structure and mass.

For the other five parameters, flat priors in the following ranges are adopted:
\begin{itemize}
\renewcommand{\labelitemi}{$\bullet$}
 \item $0<c_{200\mathrm{m}}<20$
 \item $0<e<0.9$
 \item $-\theta_\mathrm{cl}/2<\Delta_x<\theta_\mathrm{cl}/2$
 \item $-\theta_\mathrm{cl}/2<\Delta_y<\theta_\mathrm{cl}/2$
 \item $0<\theta_{e}<\pi$
\end{itemize}
The best-fitting model parameters are determined by minimizing the $\chi^2$ statistic, while constraints on individual parameters are derived by projecting the MCMC-sampled likelihood distributions onto the parameter space, with marginalization performed over the uncertainties of the other parameters. We calculated the dispersion (standard deviation) of parameters (concentration, ellipticity, and offset) from their posterior distributions of each cluster, and considered the inverse of the square of this dispersion as the weight to calculate the weighted average best-fit parameters in our subsequent analyses. 

\begin{figure}[t]
  \centering
  \begin{minipage}{0.48\columnwidth}
    \centering
    \includegraphics[width=\linewidth]{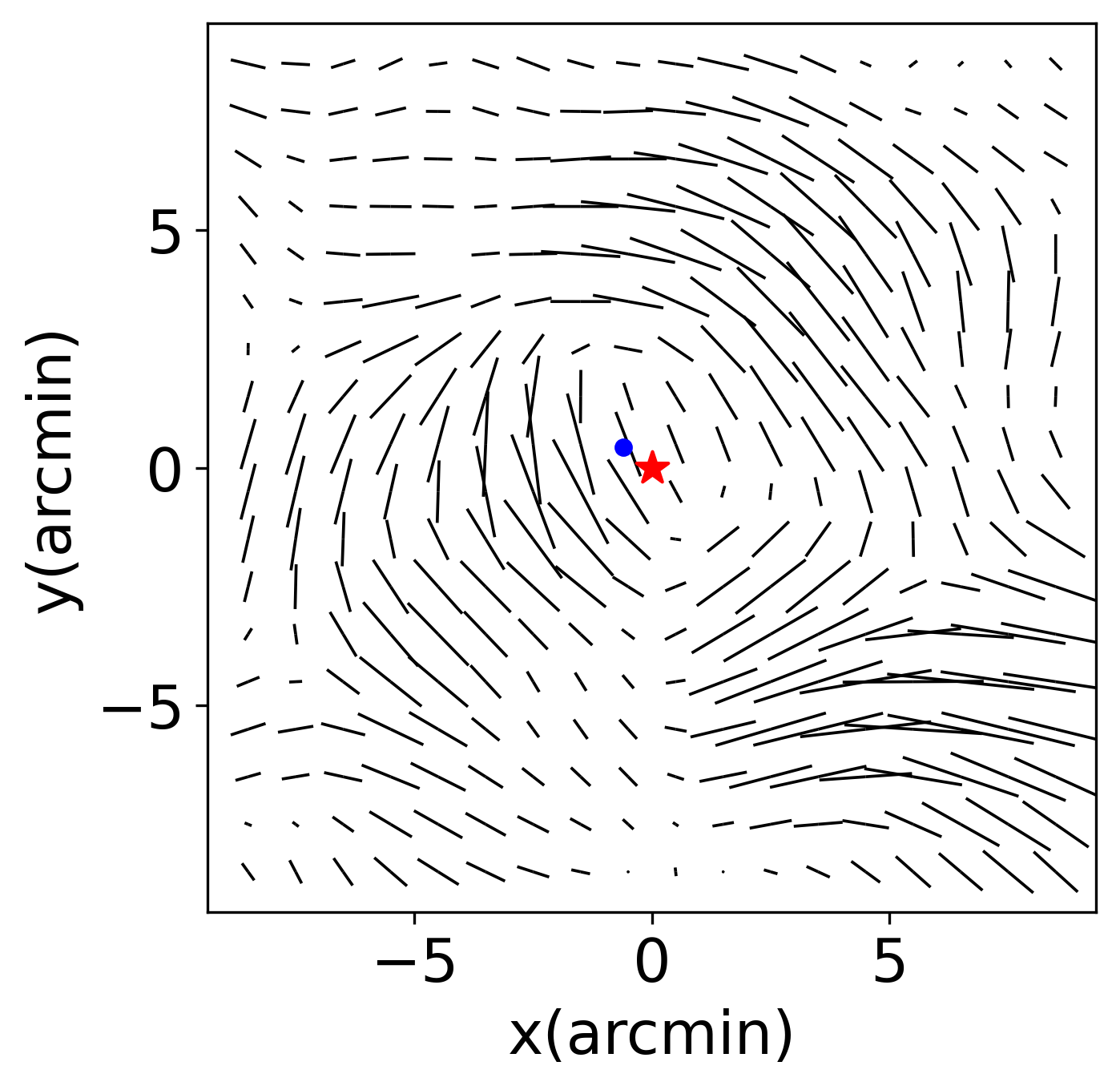}
  \end{minipage}\hfill
  \begin{minipage}{0.48\columnwidth}
    \centering
    \includegraphics[width=\linewidth]{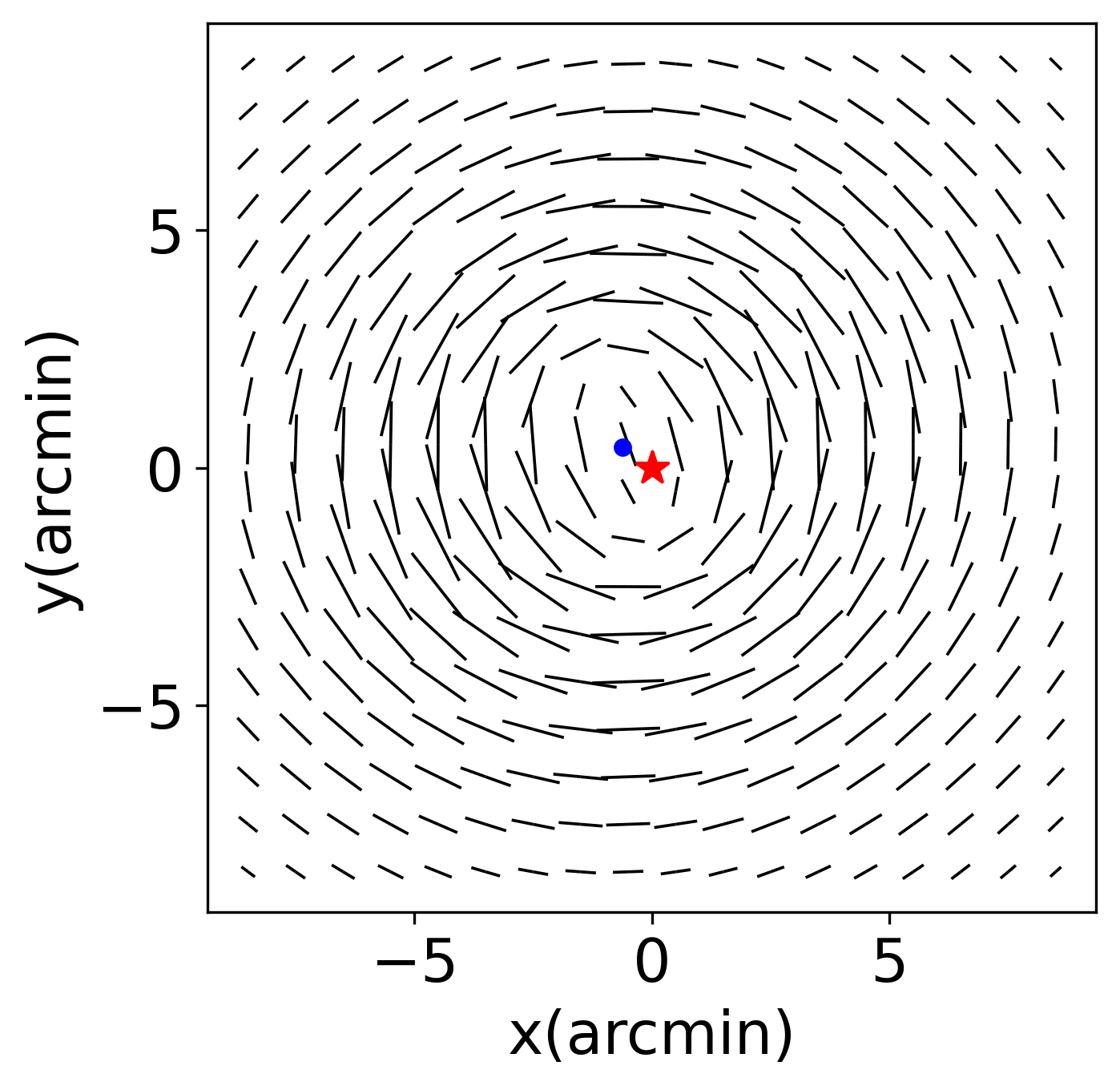}
  \end{minipage}

  \vspace{0.6em}
  \includegraphics[width=0.8\columnwidth]{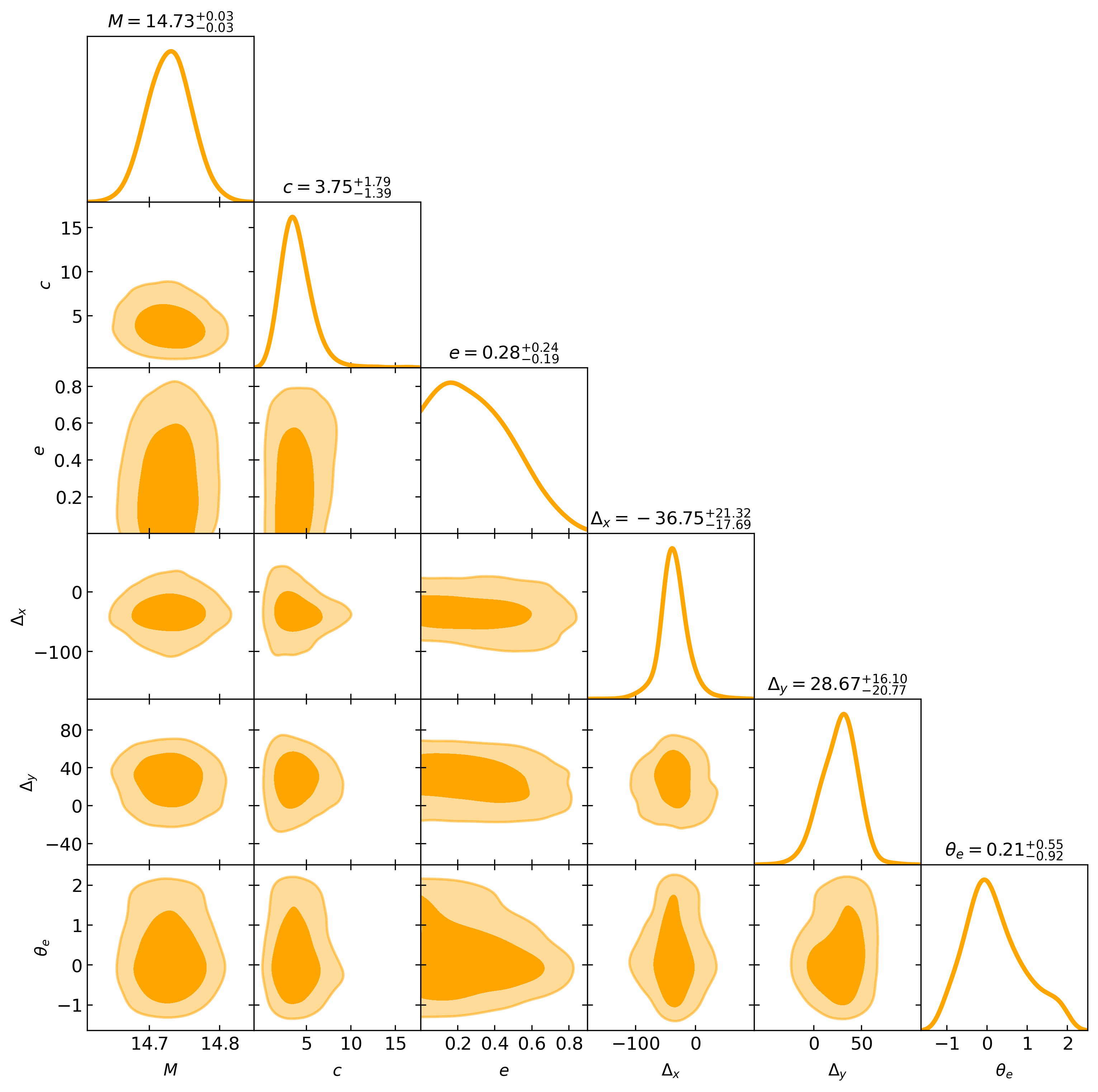}

  \caption{Upper panels: The left panel shows the observed 2D shear map for an illustrative RedMaPPer cluster ($M\simeq10^{14.73} h^{-1}M_{\odot}$), while the right panel displays the corresponding best-fitting shear field predicted by the elliptical NFW model. Each map covers an area of 18$'$$\times$18$'$. In both maps, the orientation and length of the sticks within each 1$'$$\times$1$'$ pixel indicate the local tangential direction and amplitude of the distortion, respectively. For visualization purposes, both shear fields have been smoothed with a Gaussian kernel with a full width at half maximum (FWHM) of $\approx2'$ \citep{33}. The blue dot marks the weak-lensing-derived center, and the red pentagram denotes the observed center.
  Lower panel: Posterior constraints on structural parameters of the cluster, including the halo concentration, ellipticity, and projected offsets of the mass centroid. The contours represent the $1\sigma$ and $2\sigma$ confidence levels.}
  \label{fit_results_onecluster}
\end{figure}

\section{Results} \label{sec:4}
In this section, we present our main results. The results for individual clusters are described in Section \hyperref[subsec:4.1]{\ref{subsec:4.1}}, with the observed and fitted 2D shear map and the corresponding fitting results for one RedMaPPer cluster shown as an example. The concentration-mass relation, the BCG offset distribution and the cluster ellipticity distribution of our sample are discussed in Section \hyperref[subsec:4.2]{\ref{subsec:4.2}}, Section \hyperref[subsec:4.3]{\ref{subsec:4.3}}, and Section \hyperref[subsec:4.4]{\ref{subsec:4.4}}, respectively.

\subsection{ Results for individual clusters } \label{subsec:4.1}

For our sample of 299 RedMaPPer-selected clusters, we carefully construct their 2D shear maps using the first-year HSC shear data, following the methodology described in Section \hyperref[subsec:3.1]{\ref{subsec:3.1}}. We then constrain the structural parameters of each cluster by directly fitting these shear maps with elliptical NFW models introduced in Section \hyperref[subsec:3.2]{\ref{subsec:3.2}} via MCMC (Section \hyperref[subsec:3.3]{\ref{subsec:3.3}}). The analyses primarily focus on the halo concentration, ellipticity, and the distribution of miscentering offsets. The best-fit shear maps predicted by the elliptical NFW model closely reproduce the observed shear patterns, particularly in regions where the tangential shear signal has a high signal-to-noise ratio.

The observed 2D shear map (left) and its corresponding best-fit model (right) for a RedMaPPer cluster are presented in the upper panels of Fig.\hyperref[fit_results_onecluster]{\ref{fit_results_onecluster}} as a representative example. For visualization purposes only, both the observed and fitted shear fields have been smoothed with a Gaussian kernel of full width at half maximum (FWHM) $\approx$ 2$'$ \citep{33}; please note that the unsmoothed shear fields are used in the model fitting. In this example, the size of the map is 18$'$$\times$18$'$. The orientation and length of the sticks in each 1$'$$\times$1$'$ pixel represent the tangential direction and amplitude of the distortion, respectively. The blue dot and the red pentagram represent the center determined by weak lensing and the observed center (position of the BCG), respectively. 

The lower panel of Fig.\hyperref[fit_results_onecluster]{\ref{fit_results_onecluster}} presents posterior constraints on structural parameters, including concentration, ellipticity and mass centroid offsets, of the specific cluster. The contours are for $1\sigma$ and $2\sigma$ confidence levels. For each parameter, the best-fitting value and the corresponding 68 percent confidence interval are indicated above the respective one-dimensional marginalized distribution. The projected offsets $\Delta_x$ and $\Delta_y$ are expressed in arcseconds, while the position angle $\theta_e$ is measured in radians, defined counterclockwise from the positive $x$-axis.

\begin{figure}[t]
\centering
\includegraphics[width=\linewidth, angle=0]{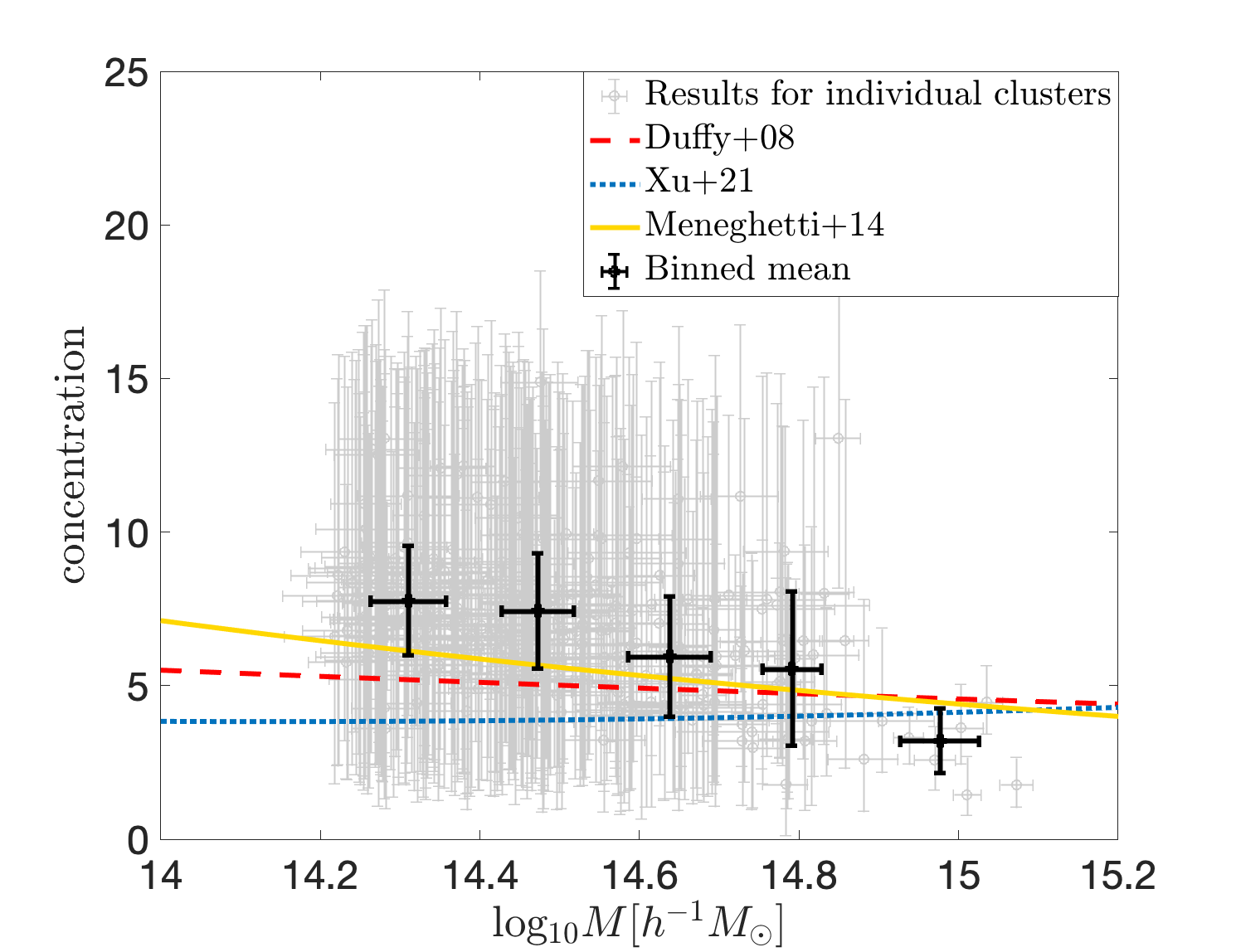}
\caption{The concentration–mass (c–M) relation. Light gray points represent the best-fitting concentration and mass for individual clusters, while black points show the weighted average values in each mass bin, with error bars indicating the standard deviation. For comparison, the results on strong-lensing selected halos from \citet{Meneghetti_SL} are shown as orange solid line. Weak-lensing-based results from \citet{19} (upturn model) are plotted as blue dot-dashed line, and the power-law model from \citet{2008Duffy} derived from dark-matter-only numerical simulations, is plotted as red dashed line, respectively.}
\label{c-M relation}
\end{figure}

\subsection{Concentration-mass relation} \label{subsec:4.2}
The concentration parameter characterizes the degree of central mass concentration within a dark matter halo or cluster. It depends on both the halo’s mass and redshift, and is closely linked to its formation and assembly history. A common observational approach to infer the average concentration is to stack lens–source pairs in bins of similar cluster mass and redshift, then fit the stacked shear profiles with parametric models \citep{19}.

Here, we measure the concentration of individual dark-matter halos and combine the results from all 299 clusters to statistically constrain the concentration-mass (c-M) relation. In Fig.\hyperref[c-M relation]{\ref{c-M relation}}, each gray point represents the best-fitting concentration and mass for a single cluster. We divided the range of $\log_{10}(M_{200\mathrm{m}})$ into five equal-width bins. Within each bin, the weighted average of the best-fit concentration and mass is computed, and the corresponding standard deviations are shown as black points with error bars.

To place our results in context, we compare the derived c–M relation with several previous studies, as shown in Fig.~\hyperref[c-M relation]{\ref{c-M relation}}. The red dashed line corresponds to the power-law model from \citet{2008Duffy}, derived from dark-matter-only numerical simulations. The blue dashed line corresponds to the upturn model proposed by \citet{19}, based on weak-lensing measurements using the DECaLS DR8 shear catalog, with a particular emphasis on the high-mass end. We also include results based on strong-lensing selected halos. \citet{Meneghetti_SL} estimate the concentrations of CLASH strong-lensing clusters through numerical simulations, using the projected mass distributions to fit NFW profiles, shown as the orange solid line.

For a fair comparison, all reference models are evaluated at the mean redshift of our sample, $z = 0.339$, and we apply appropriate conversions to ensure that their definitions of halo mass and concentration are consistent with those adopted in this work.

As shown in Fig.~\hyperref[c-M relation]{\ref{c-M relation}}, our measured c–M relation lies closer to the strong-lensing-selected results of \citet{Meneghetti_SL}, appears steeper than the power-law relation of \citet{2008Duffy}, and shows no indication of an upturn at the high-mass end. This may suggest that our sample selection and fitting methodology are more sensitive to halo concentration at the high-mass end. However, neither the upturn model derived from weak-lensing studies nor the power-law prediction obtained from dark-matter-only numerical simulations can be ruled out within the current level of observational uncertainty.  

\begin{figure}[t]
\centering
\includegraphics[width=\linewidth, angle=0]{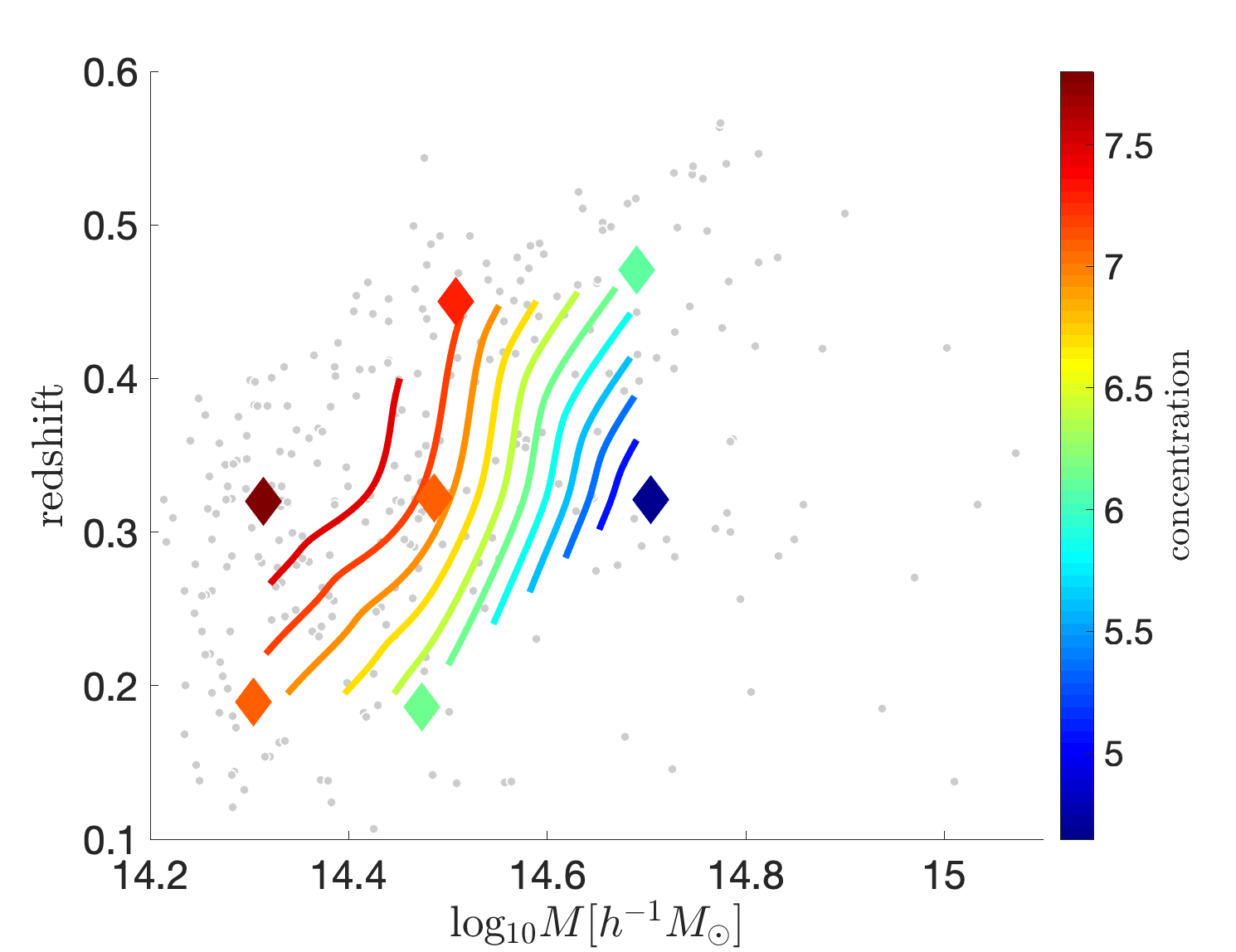}
\caption{The concentration–mass–redshift (c–M–z) relation. Gray points indicate the positions of individual clusters in the mass–redshift plane. Colored diamonds represent the weighted mean concentration within each bin. Bins containing fewer than 10 clusters are excluded from the analysis. Colored contour lines trace surfaces of constant concentration across the plane, and the color bar (labelled ``concentration") gives their numerical values.}
\label{m_z_c}
\end{figure}

We further investigate the dependence of concentration in mass-redshift space. In Fig.~\hyperref[m_z_c]{\ref{m_z_c}}, gray points denote the mass and redshift of individual clusters in our sample. We divide the sample into bins with a redshift interval of 0.15 and a $\log_{10}(M_{200\mathrm{m}})$ interval of 0.2. Bins containing fewer than 10 clusters are excluded from the analysis. Within each bin, the weighted mean concentration is computed and shown as diamond symbols. The color of each diamond reflects the average concentration value in that bin, providing a visual representation of the concentration gradient across the mass–redshift plane.

By following the direction of gradient descent along the concentration contours, we find a clear trend: halos with lower redshifts and higher masses exhibit systematically lower concentrations. This trend is consistent with the expectation that more massive clusters at lower redshift are more dynamically relaxed and have had more time to evolve toward equilibrium.

To quantitatively characterize the joint dependence of concentration on both mass and redshift, we adopt a fitting function from \citet{Meneghetti_SL} and apply it to the binned, weighted-mean concentrations in our sample:

\begin{equation}
c(M, z) = A \left( \frac{1+z_\mathrm{mean}}{1 + z} \right)^B \left( \frac{M}{M_\mathrm{mean}} \right)^C,
\label{eq:c_m_z_relation}
\end{equation}
here, $z_{\mathrm{mean}} = 0.339$ and $M_{\mathrm{mean}} = 3.058 \times 10^{14}h^{-1} M_\odot$ represent the mean redshift and the mean mass of the sample, respectively. The best-fitting parameters and their associated uncertainties corresponding to the 68 percent confidence interval are summarized in Table~\ref{para_m_z_table}.

\begin{table}[htbp]
\centering
\caption{The best-fitting parameters for the concentration–mass–redshift (c–M–z) relation derived from our cluster sample.}

\begin{tabular}{cccc}
\hline\hline
Parameter & A & B & C \\
\hline
c & $7.01^{+6.33}_{-3.98}$ & $-0.06^{+1.63}_{-1.32}$ & $-0.32^{+0.33}_{-0.32}$ \\
\hline
\end{tabular}
\label{para_m_z_table}
\end{table}
Due to limitations in sample size and observational conditions, the precision of our quantitative characterization of the c–M–z relation is somewhat constrained. Nevertheless, the overall trend reveals a negative correlation of concentration with both halo mass and redshift.

\begin{figure}[htbp]
    \centering
    \includegraphics[width=0.9\columnwidth]{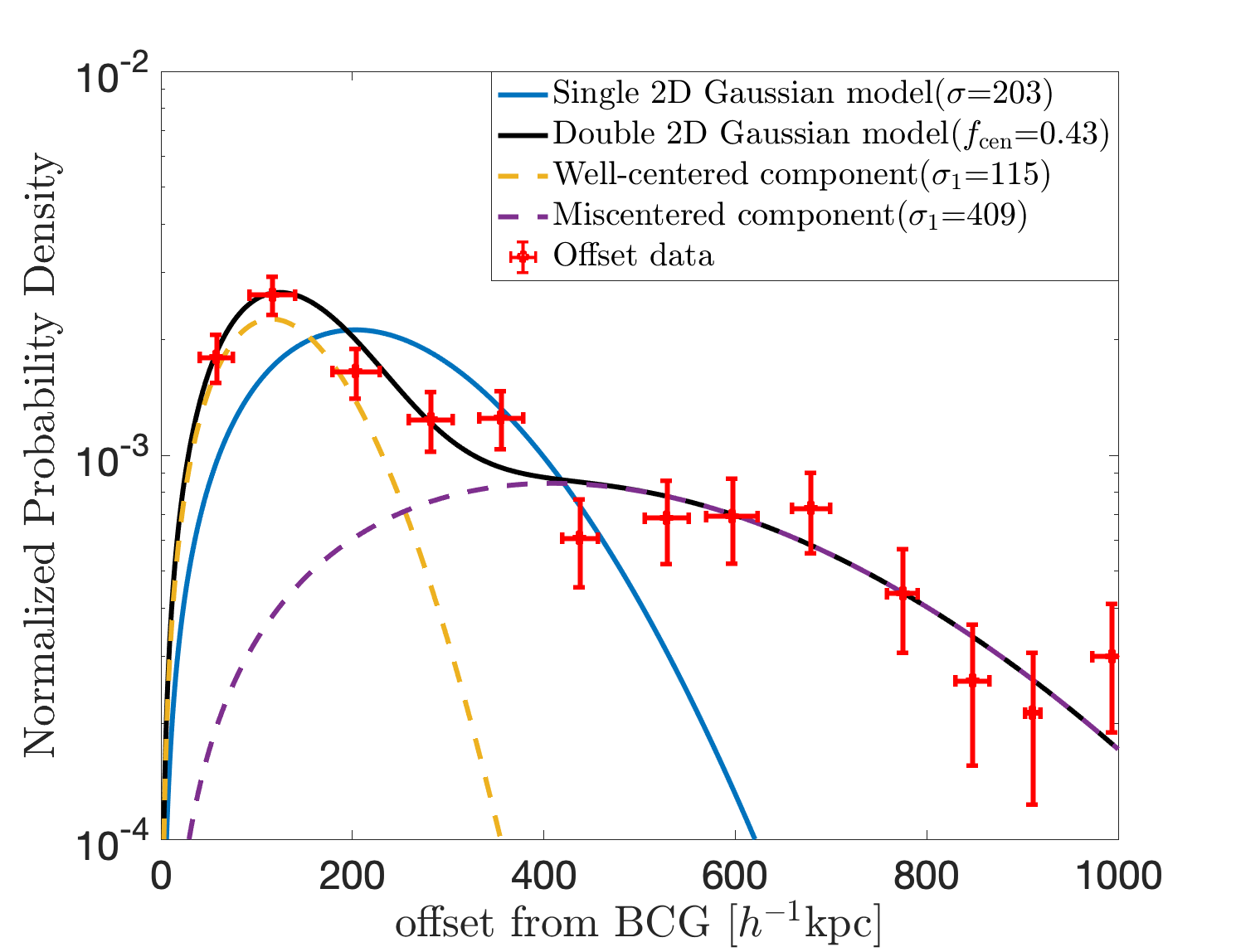}\\[0.3cm]

    \begin{subfigure}[b]{0.5\columnwidth}
        \vspace{0pt} 
        \centering
        \includegraphics[width=\columnwidth]{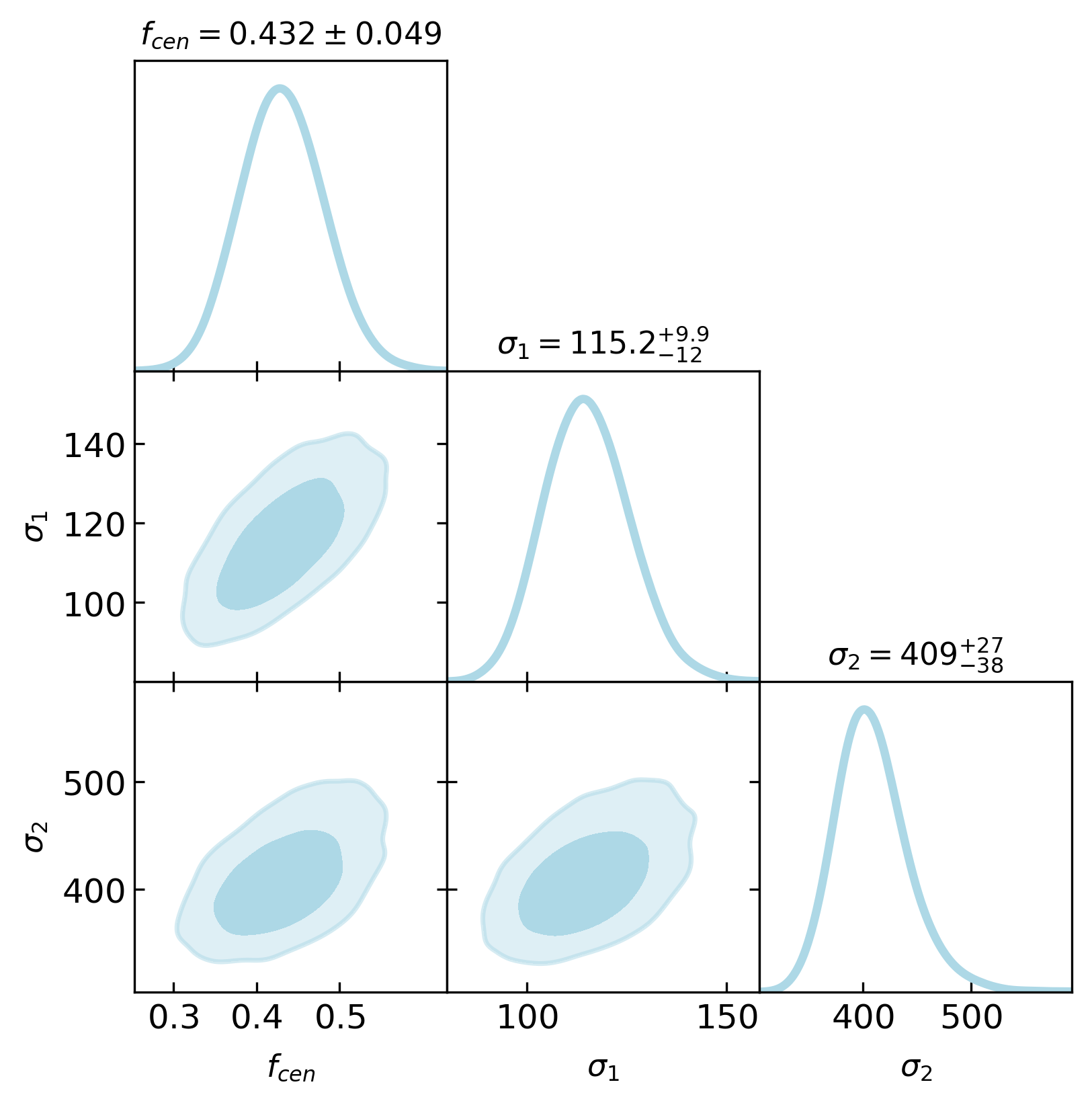} 
    \end{subfigure}%
    \hfill
    \begin{subfigure}[b]{0.45\columnwidth}
        \vspace{0pt} 
        \centering
        \includegraphics[width=\columnwidth]{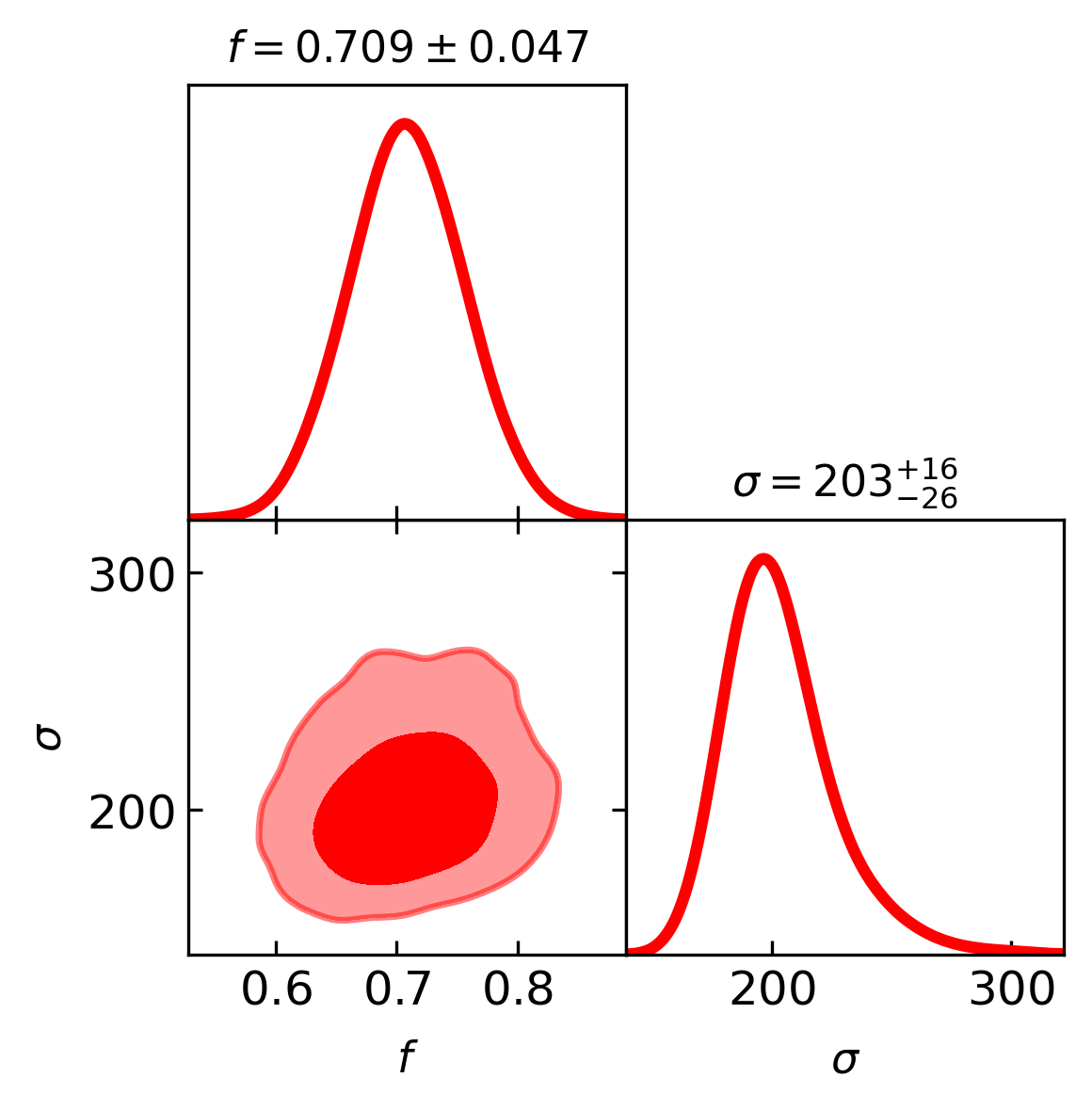}     
    \end{subfigure}

    \caption{Upper panel: Probability distribution function (PDF) of the projected offset between the BCG position and the weak-lensing-determined halo center. Red data points show the observed BCG offsets in our sample, with error bars estimated via 1000 bootstrap resamplings. The blue and black solid lines represent the best-fitting single- and two-component 2D Gaussian models, respectively. The purple and orange dashed lines correspond to the well-centered and miscentered components of the two-component model. All offset scale parameters, including $\sigma$, $\sigma_1$, and $\sigma_2$, are expressed in units of $h^{-1} \mathrm{kpc}$.
    Lower panels: Fitting results for the two-component (left) and single-component (right) 2D Gaussian models. Contours indicate $1\sigma$ and $2\sigma$ confidence levels, with best-fitting values and 68\% confidence intervals shown above each 1D marginalized distribution.}
    \label{offset_fitting}
\end{figure}

\subsection{ Offset between lensing center and BCG } \label{subsec:4.3}
In observational studies, halo centers are typically identified with either the position of a massive galaxy (e.g., the BCG) or the peak of the hot gas distribution. This is because the lensing center of mass, dominated by invisible dark matter, is not directly observable. Miscentering between the true mass center and these observational tracers is a significant source of systematic uncertainty in stacked weak-lensing analyses \citep{stack_effect,mass_bias}. Correcting for this effect not only improves the accuracy of mass measurements but also enhances our understanding of the nature of dark matter \citep{science}.

Weak lensing provides a unique and independent means of locating the lensing center of a dark matter halo, as it directly probes the projected gravitational potential and is sensitive to the full matter distribution. It thus offers a complementary perspective to optical indicators (e.g., BCGs) and thermal signals from the Sunyaev–Zel’dovich (SZ) effect \citep{miscentering-SZ}.

In modeling the miscentering effect in weak lensing analyses, a two-dimensional (2D) Gaussian distribution is widely adopted as a realistic representation of offset distributions. Following the approach of \citet{oguri_2018}, we model the BCG offset distribution using a two-component 2D Gaussian function, as shown in Fig.~\hyperref[offset_fitting]{\ref{offset_fitting}}, which enables a flexible description of both well-centered and miscentered populations:

\begin{equation}
\begin{aligned}
    P(R_\mathrm{off})= f_\mathrm{cen}\frac{R_\mathrm{off}}{\sigma^2_1} \exp(-R_\mathrm{off}^2/2\sigma^2_1) \\ 
  + (1-f_\mathrm{cen})\frac{R_\mathrm{off}}{\sigma^2_2}\exp(- R_\mathrm{off}^2/2\sigma^2_2),
\end{aligned}
\label{2DGaussian}
\end{equation}
here, $R_{\mathrm{off}}$ denotes the projected separation between the weak-lensing-determined halo center and the BCG position. The parameter $f_{\mathrm{cen}}$ represents the fraction of well-centered clusters, while $\sigma_1$ and $\sigma_2$ describe the characteristic offset scales of the well-centered and miscentered components, respectively. Parameter uncertainties are estimated via 1000 bootstrap resamplings. The reduced $\chi^2$ values for the single- and two-component models are 8.58 and 0.87, respectively, confirming the superior fit of the two-component model.

The upper panel of Fig.~\hyperref[offset_fitting]{\ref{offset_fitting}} displays the weighted distribution of BCG–WL centroid offsets. The blue solid line represents a single 2D Gaussian fit with a characteristic scale of $\sigma = 203^{+16}_{-26}~h^{-1} \mathrm{kpc}$, which evidently fails to capture the bimodality. The best-fitting parameters for the two-component model are $f_{\mathrm{cen}} = 0.43 \pm 0.05$, $\sigma_1 = 115.2 \pm ^{+9.9}_{-12}~h^{-1} \mathrm{kpc}$ for the well-centered population, and $\sigma_2 = 409^{+27}_{-38} ~h^{-1} \mathrm{kpc}$ for the miscentered population.

To test the robustness of our offset measurements, we noticed that for a few clusters, the inferred offsets were close to the edge of the default fitting range $[-\theta_\mathrm{cl}/2, \theta_\mathrm{cl}/2]$. To evaluate whether this boundary could bias the results, we repeated the fitting for these clusters using an expanded data range of $[-\theta_\mathrm{cl}, \theta_\mathrm{cl}]$, while keeping the offset prior unchanged. The resulting parameter constraints remain consistent within uncertainties, confirming the stability and reliability of the fitted offset distribution.

We further divide the sample into high- and low-mass subsamples based on the median halo mass. Offsets in the low-mass bin exhibit larger values of $\sigma_1$ and $\sigma_2$, suggesting that lower-mass halos in our sample may be more affected by surrounding large-scale structure and projection effects, leading to greater displacements between the WL center and the BCG.
\begin{figure}[t]
\centering
\includegraphics[width=\linewidth, angle=0]{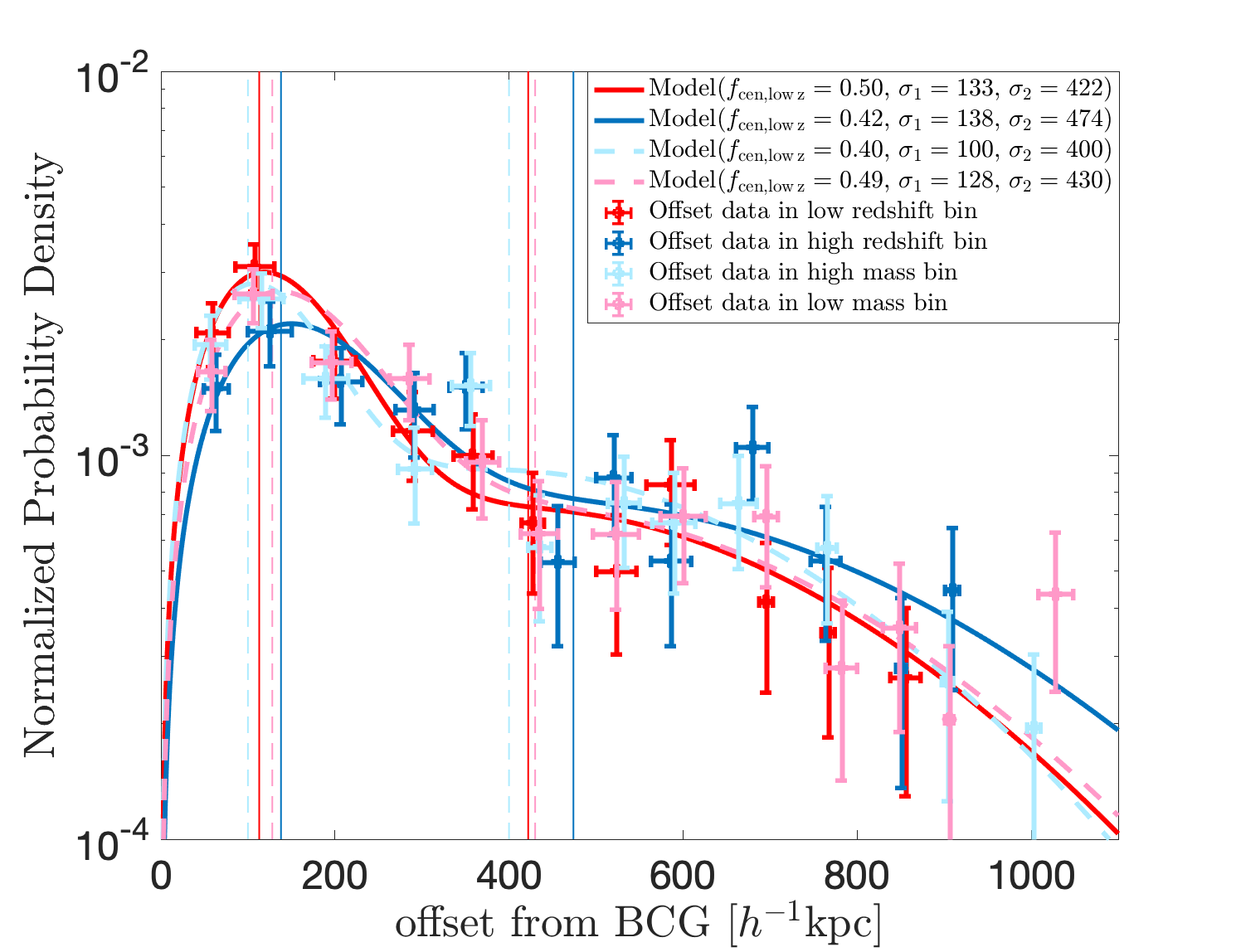}
\caption{Probability distribution functions (PDFs) of the projected offsets between the BCG and the weak-lensing-determined halo center for different redshift and mass bins. All error bars are estimated from 1000 bootstrap resamplings. Red and blue points represent the low- and high-redshift subsamples, respectively, with solid lines showing the best-fitting two-component 2D Gaussian models. Corresponding vertical lines indicate the values of $\sigma_1$ and $\sigma_2$. Pink and light blue points show the low- and high-mass subsamples, respectively, with dashed lines and vertical markers indicating the best fits and offset scales.}
\label{dif_bin_offset}
\end{figure}

Similarly, when dividing the sample into high- and low-redshift bins, we find that the low-redshift clusters tend to have higher values of $f_{\mathrm{cen}}$ and smaller offset scales. This trend is consistent with the expectation that halos become more dynamically relaxed over time, resulting in better alignment between the mass centroid and the BCG.

These comparisons are illustrated in Fig.~\hyperref[dif_bin_offset]{\ref{dif_bin_offset}}, which shows the projected offset distributions and the corresponding best-fit parameters for different redshift and mass bins.

We noticed that the miscentering parameters inferred from our BCG-WL offsets differ from those reported in previous studies. For example, the $f_{\mathrm{cen}}$ value shown in Fig.~\hyperref[offset_fitting]{\ref{offset_fitting}} is lower than that obtained by \citet{oguri_2018}, while our offset scales ($\sigma_1$ and $\sigma_2$) are larger than their reported values of $46 \pm 9 ~ h^{-1} \mathrm{kpc}$ and $260 \pm 40 ~ h^{-1} \mathrm{kpc}$, respectively. This discrepancy may arise from differences in center definitions and sample selection. Notably, \citet{oguri_2018} measured the offsets between BCGs and X-ray centers of clusters, with the X-ray temperatures estimated using a mass–temperature relation based on $M_{500}$, which corresponds to more compact regions than the $M_{200\mathrm{m}}$ used in our analysis.
\begin{figure}[t]
\centering
\includegraphics[width=\linewidth, angle=0]{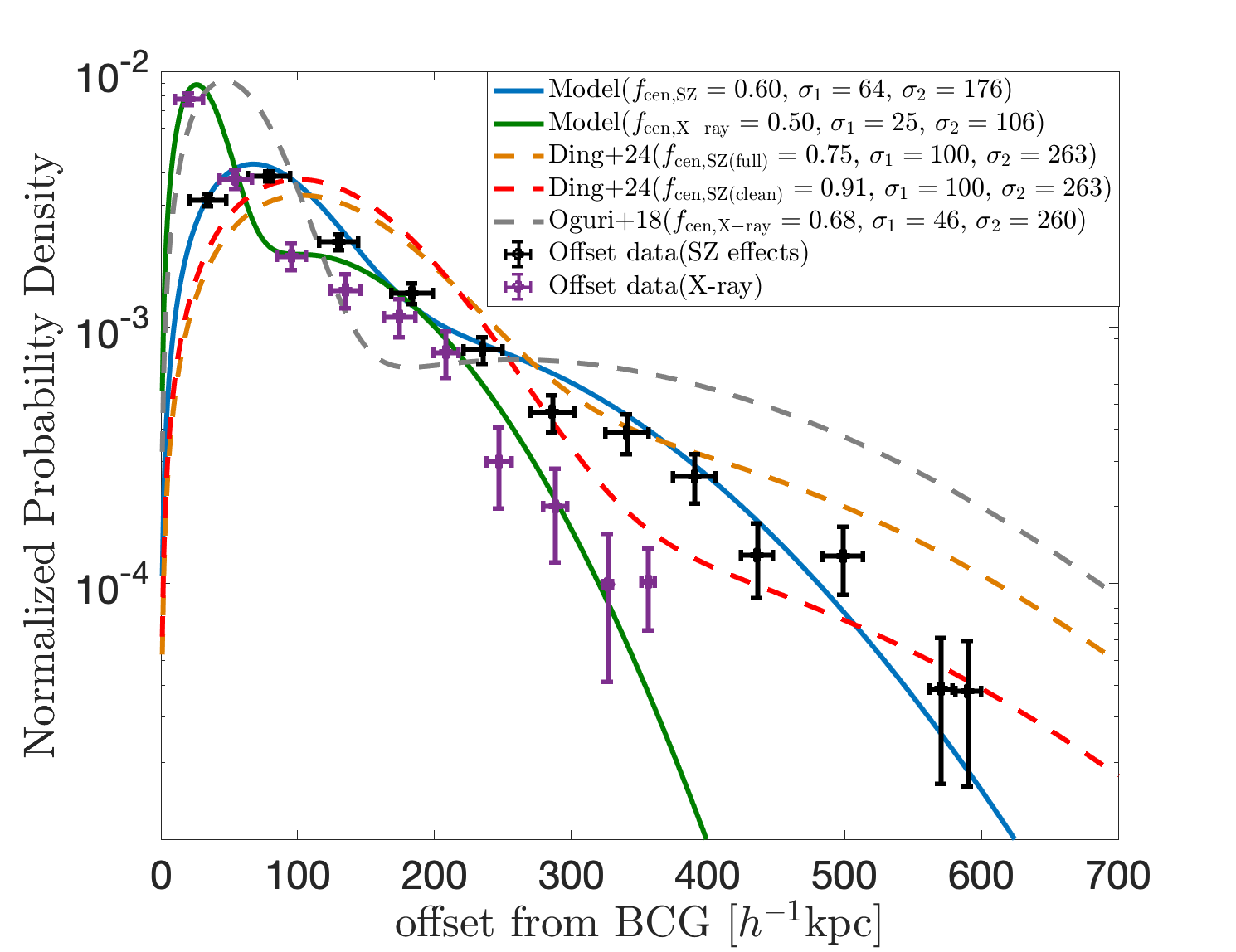}
\caption{Probability distribution functions (PDFs) of the projected offsets between the BCG position and the SZ- or X-ray-determined cluster centers. Black and purple data points represent the offsets between the RedMaPPer optical centers and the SZ (ACT DR5) or X-ray cluster centers, respectively. The blue and green solid lines show the best-fitting two-component 2D Gaussian models for the SZ and X-ray cases. The orange and red dashed lines show results from \citet{miscentering-SZ}: the orange line corresponds to all 186 optical–SZ cross-matched clusters, while the red line excludes clusters affected by astrophysical effects (e.g., ongoing mergers) or systematic biases in the HSC data and cluster-finding algorithm. The gray dashed line shows the X-ray–optical offset distribution from \citet{oguri_2018}, based on HSC Wide S16A clusters. All offsets are shown in physical units of $h^{-1}\mathrm{kpc}$.}
\label{SZ_xray_offset}
\end{figure}
Motivated by the relatively large WL–BCG offsets, we further investigate whether this trend reflects genuine physical differences between weak-lensing and baryon-based tracers. As a first step, we compare the optical centers from RedMaPPer with those from two other optical catalogs - HSC S19A \citep{HSCSSPDR3} and WH2022 \citep{WH2022}. For each HSC or WH2022 cluster, we identify the nearest RedMaPPer counterpart using the \texttt{match\_to\_catalog\_sky} method in \texttt{astropy}'s \texttt{SkyCoord} class \citep{Astropy}, with a maximum separation of 1 arcminute. We find that more than 85\% of the matched clusters have projected offsets smaller than $100\mathrm{kpc}$, indicating that the optical-optical center differences are negligible.

We then compare the RedMaPPer optical centers with those determined from SZ observations (ACT DR5; \citealt{ACT}) and X-ray data \citep{eROSITA}, computing the projected physical offsets using RedMaPPer photometric redshifts. In this calculation, we adopt a maximum offset threshold corresponding to a typical cluster size $(\sim1 h^{-1}\mathrm{Mpc})$ to exclude spurious large separations. The resulting offset distributions and their best-fitting two-component models are shown in Fig.~\hyperref[SZ_xray_offset]{\ref{SZ_xray_offset}}, along with several results from the literature \citep{miscentering-SZ, oguri_2018} for comparison. Although the exact values of $f_{\mathrm{cen}}$, $\sigma_1$, and $\sigma_2$ vary among studies, the overall shapes of the offset distributions are broadly consistent.

Therefore, the offsets involving SZ or X-ray centers are significantly smaller than those based on weak-lensing-determined centers. This suggests that weak lensing is identifying a different characteristic center—one that traces the projected gravitational potential and is influenced by the full matter distribution along the line of sight weighted by lensing efficiency.

Unlike X-ray or SZ signals, which trace the distribution of hot baryonic gas through electron density and temperature, weak lensing directly probes the total mass, including both baryonic and dark matter, and is independent of the dynamical state of the cluster. Consequently, in unrelaxed systems or in clusters embedded in complex environments with substantial line-of-sight structures, weak-lensing-derived centers may appear significantly offset from the baryon-based tracers. These intrinsic differences in physical sensitivity and projection response among tracers likely contribute to the broader offset distributions observed in WL-based analyses.

For clusters with the largest centroid offsets, the projected distributions of member galaxies are found to be frequently bimodal, suggesting ongoing mergers or significant dynamical disturbances; however, given the current data quality and the unknown dynamical information of the clusters, we refrain from definitive classification and defer a more detailed analysis to future work.

\subsection{ Distribution of halo ellipticities } \label{subsec:4.4}
Ellipticity is a key parameter used to characterize the projected shape of dark matter halos along the line of sight, and its importance in understanding halo structure is well established. Although the spherically symmetric NFW model successfully reproduces the spherical average of the halo density profile, both $N$-body simulations \citep{2} and observations \citep{33,halo-shear-shear} consistently show that halos are intrinsically non-spherical.

Moreover, halo shape provides insights into the nature of dark matter. For instance, halos in self-interacting dark matter (SIDM) models are predicted to be more spherical than those in collisionless cold dark matter (CDM) scenarios, as particle self-interactions tend to isotropize the dark matter distribution \citep{SIDMshape}.

\begin{figure}[htbp]
    \centering
    \begin{subfigure}[b]{0.5\textwidth}
        \includegraphics[width=1\textwidth]{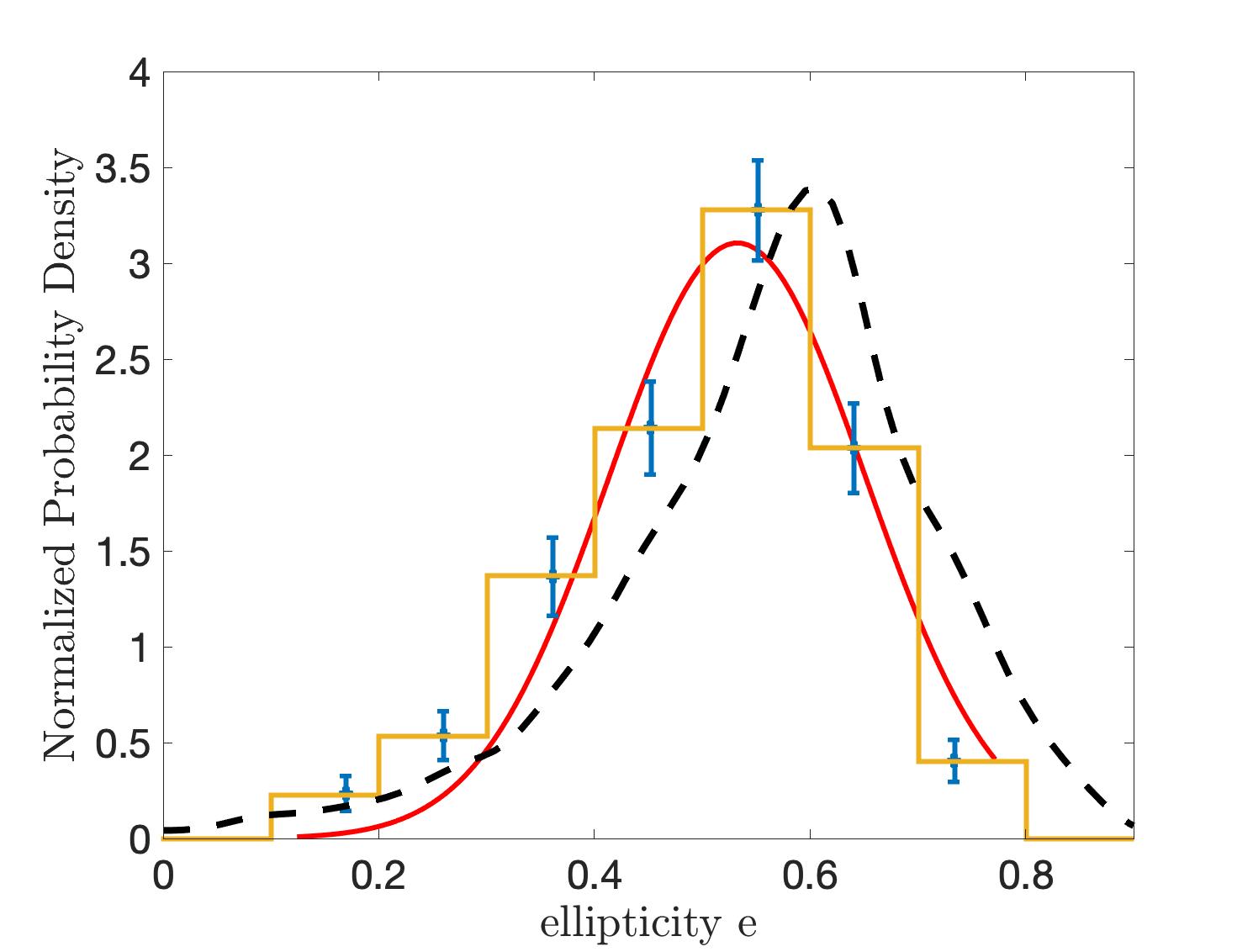}
        \label{fig:image1}
    \end{subfigure}
    
    \vspace{1mm} 
    
    \begin{subfigure}[b]{0.5\textwidth}
        \centering
        \includegraphics[width=1\textwidth]{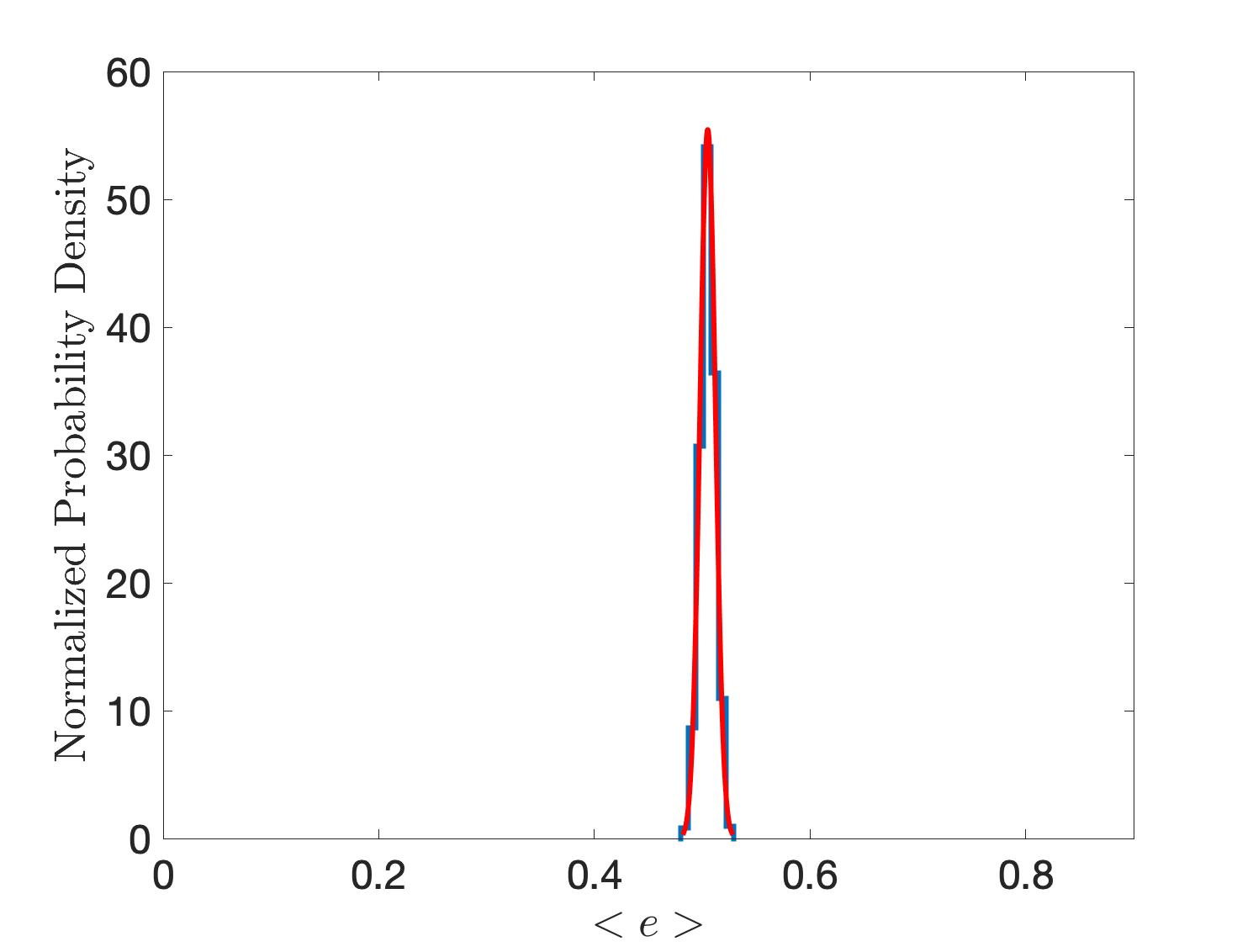}
        \label{fig:image2}
        \end{subfigure}
\caption{ Upper panel: 
Probability distribution function (PDF) of the ellipticity values derived from individual shear map fittings of each cluster. Blue points represent the observed distribution, with error bars estimated from 1000 bootstrap resamplings. The red curve shows a Gaussian fit to the distribution, centered at $e = 0.530 \pm 0.168$. For comparison, the black dashed line represents the ellipticity distribution reported by \citet{halo-shear-shear}, converted to match our ellipticity definition.
Lower panel: 
PDF of the sample-averaged ellipticity, $\langle e \rangle$, computed from the full cluster ensemble. The red curve shows a Gaussian fit centered at $\langle e \rangle = 0.505 \pm 0.007$.
}
\label{PDF_e}
\end{figure}

In our analysis, we perform an individual shear map fitting for each cluster to estimate its own ellipticity, and then combine the resulting values to construct the overall probability distribution, as shown in Fig.~\hyperref[PDF_e]{\ref{PDF_e}}. This approach captures the intrinsic variation of halo shapes across the sample. Errors are estimated using 1000 bootstrap resamplings. The measured ellipticity for the full sample is $e = 0.530 \pm 0.168$, and the mean ellipticity is $\langle e \rangle = 0.505 \pm 0.007$, which is in good agreement with predictions from $N$-body simulations \citep{2} and previous observational results \citep{halo-shear-shear}.

\begin{table}[htbp]
\centering
\caption{The ellipticity distribution in different mass–redshift bins.}
\begin{tabular}{l|l l}
\hline\hline
Samples & Low z & High z  \\
\hline
Low $M_{200m}$ & $0.480\pm 0.12$ & $0.510\pm 0.09$   \\
High $M_{200m}$ & $0.465\pm 0.17$ & $0.492\pm 0.13$  \\
\hline
\end{tabular}
\tablefoot{
The sample is split into four bins by mass and redshift, divided at the mean mass and redshift values: low-/high-mass and low-/high-redshift.
}
\label{e_m_z_table}
\end{table}

In Table~\ref{e_m_z_table}, we divide the sample into four subsets in the mass–redshift plane based on the mean halo mass and mean redshift. Specifically, clusters are grouped into quadrants: (1) low mass \& low redshift, (2) low mass \& high redshift, (3) high mass \& low redshift, and (4) high mass \& high redshift. We compute the average ellipticity in each bin.
The results show that halos with higher mass and lower redshift tend to exhibit slightly lower ellipticities, implying that they appear more spherical along the line of sight. This trend contrasts with predictions from dark-matter-only cosmological simulations \citep{2,3}, which include only gravitational interactions. These simulations typically find that more massive and lower-redshift halos are slightly more elliptical due to persistent anisotropic mass accretion from the surrounding cosmic web. Interestingly, our findings align more closely with the trend for halo masses larger than $10^{13.5}~h^{-1}M_{\odot}$ in the hydrodynamical IllustrisTNG simulation \citep{halo-shear-shear}, which incorporates baryonic physics.  In this context, more massive halos are expected to assemble earlier and more rapidly, reducing subsequent interactions with their large-scale environments and thus evolving into more isotropic and spherical configurations.

\section{Conclusion and discussion} \label{sec:5}
In this paper, we analyze 299 individual RedMaPPer-selected clusters by constructing their 2D shear maps from the HSC first-year shear catalog with high data quality. We statistically investigate key halo properties, including concentration, ellipticity, and the BCG offset. To mitigate the impact of intrinsic mass uncertainties on the statistical concentration–mass (c–M) relation, we adopt a mass–richness prior from \citet{mass-richness_equation} in our fitting procedure. Furthermore, we use the inverse of the posterior variance of each fitted parameter as the statistical weight when computing weighted average values. Our main conclusions are as follows:

\begin{itemize}
\renewcommand{\labelitemi}{$\bullet$}
\item We find that halo concentration decreases with increasing mass in our sample. The derived c–M relation is steeper than those obtained from weak lensing analyses but more consistent with trends seen in strong-lensing-selected cluster studies. This may reflect that our sample selection and fitting method are more sensitive to halo concentration at the high-mass end. In the mass–redshift plane, we observe that halos with lower redshift and higher mass tend to have lower concentrations.

\item The distribution of offsets between the observed optical center and the weak-lensing-determined center is well described by a two-component 2D Gaussian model, as defined in Eq.~\hyperref[2DGaussian]{\ref{2DGaussian}}. The characteristic offset scales are $\sigma_1 = 115.2 ^{+9.9}_{-12}~h^{-1} \mathrm{kpc}$ and $\sigma_2 = 409^{+27}_{-38}~h^{-1} \mathrm{kpc}$, corresponding to the well-centered and miscentered populations, respectively. The fraction of well-centered clusters is measured to be $f_{\mathrm{cen}} = 0.43 \pm 0.05$. We find that halos in the lower mass bin tend to exhibit larger offsets, possibly due to their greater susceptibility to perturbations from surrounding large-scale structures and enhanced lensing projection effects. In comparison, lower-redshift clusters exhibit smaller offsets and higher centering fractions, consistent with the expectation that clusters become more relaxed over cosmic time. The weak-lensing-based offsets are systematically larger than those involving X-ray or SZ-defined centers \citep{oguri_2018, miscentering-SZ}, likely reflecting differences in tracer sensitivity and center definitions. In particular, weak lensing probes the total projected mass distribution—including dark matter—while X-ray and SZ observations trace the hot baryonic gas, which is more tightly correlated with the optical BCG position. These intrinsic physical differences among tracers naturally lead to broader offset distributions in WL-based measurements. Our results demonstrate that modeling miscentering with a flexible two-component model significantly improves the reliability of halo property inference. This is crucial for reducing systematics in galaxy–galaxy lensing analyses and for ensuring the robustness of high-precision cosmological probes, such as cluster abundance studies \citep{eROSITA} and three-by-two-point (3×2pt) analyses \citep{KiDs_3*2pt,HSC_3*2pt,DEC_3*2pt}, where even small biases in halo centering can impact cosmological parameter constraints. A more detailed treatment of these effects will be presented in a forthcoming paper.

\item The ellipticity distribution derived from individual cluster fittings has a mean value of $e = 0.530 \pm 0.168$, and the sample-averaged ellipticity is $\langle e \rangle = 0.505 \pm 0.007$. In the mass–redshift plane, halos with higher mass and lower redshift appear more spherical, exhibiting slightly lower average ellipticities. This trend is opposite to predictions from dark-matter-only $N$-body simulations but is more closely consistent with the trend for halo masses larger than $10^{13.5}~h^{-1}M_{\odot}$ from the IllustrisTNG simulations \citep{halo-shear-shear} that include baryonic effects. In those models, baryonic cooling contracts matter toward the center, increasing halo roundness \citep{Lau2012}, while isotropic momentum redistribution from merger events also promotes spherical shapes. A similar effect is observed in self-interacting dark matter models, where particle collisions isotropize the halo structure \citep{SIDMshape}. Although our current measurements cannot distinguish between these two physical mechanisms, the results offer meaningful constraints on the nature of dark matter.

\end{itemize}
Overall, our results demonstrate that 2D weak-lensing analysis is a powerful tool for investigating the structural properties of individual galaxy clusters, and it holds significant potential for improving the treatment of systematic uncertainties in traditional stacking methods. Future wide-field surveys, such as the Vera C. Rubin Observatory, the Euclid mission, and CSST, will deliver larger samples of clusters and background galaxies, enabling more precise constraints on halo properties and offering new opportunities to probe the fundamental nature of dark matter.

\begin{acknowledgements}

This research is supported by National Key R\&D Program of China No. 2022YFF0503403. X.K.L. acknowledges the support from NSFC grant No. 12173033, the grants from the China Manned Space Projects with No. CMS-CSST-2021-B01, the ``Yunnan Key Laboratory of Survey Science" with project No. 202449CE340002 and a ``Yunnan Provincial Top Team Projects” with project No. 202305AT350002. H.Y.S. acknowledges the support from the Ministry of Science and Technology of China (grant No. 2020SKA0110100), Key Research Program of Frontier Sciences, CAS, Grant No. ZDBS-LY-7013 and Program of Shanghai Academic/Technology Research Leader. Z.H.F. acknowledges the support from NSFC grant No. 11933002, and the grant
from the China Manned Space Projects with No. CMS-CSST-2021-A01. We also acknowledge the support from the science research grants from the China Manned Space Project with No. CMS-CSST-2025-A03 and CMS-CSST-2025-A05. Part of the computations in this study were carried out on the servers of the SWIFAR Cosmology Group and the Yunnan University Astronomy Supercomputer.

The Hyper Suprime-Cam (HSC) collaboration includes the astronomical communities of Japan and Taiwan, and Princeton University. The HSC instrumentation and software were developed by the National Astronomical Observatory of Japan (NAOJ), the Kavli Institute for the Physics and Mathematics of the Universe (Kavli IPMU), the University of Tokyo, the High Energy Accelerator Research Organization (KEK), the Academia Sinica Institute for Astronomy and Astrophysics in Taiwan (ASIAA), and Princeton University. Funding was contributed by the FIRST program from the Japanese Cabinet Office, the Ministry of Education, Culture, Sports, Science and Technology (MEXT), the Japan Society for the Promotion of Science (JSPS), Japan Science and Technology Agency (JST), the Toray Science Foundation, NAOJ, Kavli IPMU, KEK, ASIAA, and Princeton University. 

This paper makes use of software developed for Vera C. Rubin Observatory. We thank the Rubin Observatory for making their code available as free software at http://pipelines.lsst.io/.

This paper is based on data collected at the Subaru Telescope and retrieved from the HSC data archive system, which is operated by the Subaru Telescope and Astronomy Data Center (ADC) at NAOJ. Data analysis was in part carried out with the cooperation of Center for Computational Astrophysics (CfCA), NAOJ. We are honored and grateful for the opportunity of observing the Universe from Maunakea, which has the cultural, historical and natural significance in Hawaii.

The Pan-STARRS1 Surveys (PS1) and the PS1 public science archive have been made possible through contributions by the Institute for Astronomy, the University of Hawaii, the Pan-STARRS Project Office, the Max Planck Society and its participating institutes, the Max Planck Institute for Astronomy, Heidelberg, and the Max Planck Institute for Extraterrestrial Physics, Garching, The Johns Hopkins University, Durham University, the University of Edinburgh, the Queen’s University Belfast, the Harvard-Smithsonian Center for Astrophysics, the Las Cumbres Observatory Global Telescope Network Incorporated, the National Central University of Taiwan, the Space Telescope Science Institute, the National Aeronautics and Space Administration under grant No. NNX08AR22G issued through the Planetary Science Division of the NASA Science Mission Directorate, the National Science Foundation grant No. AST-1238877, the University of Maryland, Eotvos Lorand University (ELTE), the Los Alamos National Laboratory, and the Gordon and Betty Moore Foundation.

\end{acknowledgements}

\bibliography{paper}{}

\bibliographystyle{aa}

\end{document}